\documentclass[a4paper,12pt] {JHEP3}  
\usepackage{epsfig,multicol,cite}
\def\be{\begin{equation}}
\def\bea{\begin{eqnarray}}
\def\ee{\end{equation}}
\def\eea{\end{eqnarray}}
                              \def\barr{\begin{array}}
                              \def\earr{\end{array}}
                              \def\gev{\: \rm GeV} 
                               
\def\dis{\displaystyle}
\def\rp{R\!\!\!/ _p}

\def\Etmiss{$E_{T}^{miss}$}

\def\m0{$m_{0}$}

\def\l {\lambda }

\def \Eslash {E \kern-.9em\slash }
\def \pslash {p \kern-.5em\slash }
\def \kslash {k \kern-.5em\slash }

\def\dofig#1#2{\epsfig{file=#2,width=#1}}
\def\dofigs#1#2#3{\epsfig{file=#2,width=#1}\epsfig{file=#3,width=#1}}
\newcommand{\newc}{\newcommand}
\newc{\beq}{\begin{eqnarray}}
\newc{\eeq}{\end{eqnarray}}
\newc{\dqu}{\delta_{qu}}
\newc{\dqd}{\delta_{qd}}
\newc{\non}{\nonumber}
\newc{\noi}{\noindent}

\catcode`@=11 
\def \gsim{\mathrel{\mathpalette\@versim>}}
\def \lsim{\mathrel{\mathpalette\@versim<}}
\def \@versim#1#2{\lower0.4ex\vbox{\baselineskip\z@skip\lineskip\z@skip
     \lineskiplimit\z@\ialign{$\m@th#1\hfil##\hfil$%
     \crcr#2\crcr\sim\crcr}}}
\catcode`@=12 
\preprint{MRI-P-020702\\
CERN-TH/2002-142\\
IISC-CTS-09/02\\
ATL-PHYS-2002-016\\[2ex]
{\large \tt hep-ph/0207248}}
\title{Measuring $R$-parity-violating couplings 
in dilepton production at the LHC}
\author{Debajyoti Choudhury\footnote{debchou@mri.ernet.in}\\
HarishChandra Research Institute, Chhatnag Road, 
Jhusi, Allahabad 211 019, India. 
}
\author{Rohini M. Godbole\footnote{Rohini.Godbole@cern.ch, 
    rohini@cts.iisc.ernet.in}\\
Theory Division, CERN, Geneva 23, Switzerland.\\
and\\
Centre for Theoretical Studies, 
Indian Institute of Science, Bangalore 560 012, India}
\author{Giacomo Polesello\footnote{Giacomo.Polesello@cern.ch}\\
INFN, Sezione di Pavia, Via Bassi 6, Pavia 27100, Italy
}
\abstract
{We revisit the issue of probing $R$-violating couplings 
of  supersymmetric 
theories at hadronic colliders, particularly at the LHC.
Concentrating on dimuon production, 
an evaluation of the optimal sensitivity to the $R$-violating 
coupling is performed through a maximum likelihood analysis.
The measurement uncertainties are evaluated through
a study of fully generated events processed through a fast simulation
of the ATLAS detector.  It is found that a host of $R$-violating couplings 
can be measured to a statistical accuracy of better than $10\%$, over a 
significant part of the $m_{\tilde f}$ -- $\lambda$ parameter space still 
allowed by  low energy measurements.  Since the bounds thus obtained 
do not simply scale as the squark mass, one can do significantly better at 
the LHC than at the Tevatron.  The same analysis can also be extended to 
assess the reach of the LHC to effects due to any non-SM structure of the 
four-fermion amplitude, caused by exchanges of new particles with different 
spins such as  leptoquarks and  gravitons  that are suggested by various 
theoretical ideas.}
\begin{document}
\section[Introduction]{Introduction} 
%
%
%
Within the Standard Model (SM), electroweak gauge invariance 
ensures the conservation of both lepton number and baryon number, at 
least in the perturbative context. However, this is not so within the
Minimal Supersymmetric Standard Model (MSSM).
The most general superpotential respecting the gauge symmetries of the 
SM contains bilinear and trilinear terms which do not 
conserve either the baryon number ($B$) or the  lepton number ($L$).  
Clearly, the simultaneous presence of both lepton- and baryon-number-violating 
operators could lead to very rapid proton decay, especially for TeV scale 
sparticle masses. The existence of all such terms can be forbidden by 
postulating a discrete symmetry~\cite{Fayet:1977yc,Farrar:1978xj},
called $R$-parity, which implies a conserved quantum number
$R_p \equiv (-1)^{3B + L +S}$, where $S$ stands for the spin of the
particle. The very definition implies that all the SM particles
have $R_p = +1$, while all the superpartners are odd under this symmetry. 
Thus, apart from suppressing proton decay, it also guarantees 
the stability of the lightest supersymmetric particle (LSP), thereby
offering a ready-made candidate for cold dark matter.  

However, while a conserved $R$-parity seems desirable, 
it is perhaps too strong a requirement to be imposed. 
For one, the measure is an {\em ad hoc} one and there does not exist an 
overriding theoretical motivation for imposing this symmetry, 
especially since a suppression of the proton decay rate 
could as well be achieved by ensuring that one of $B$ and $L$ is conserved.
Indeed, it has been argued~\cite{Ibanez:1992pr} that 
this goal is better served by imposing a generalized 
baryon parity instead. Unlike $R$-parity, this latter 
($Z_3$) symmetry also serves to eliminate dimension-5 operators that 
could potentially have led to proton decay. Furthermore, non-zero $\rp$ 
couplings provide a means of generating the small neutrino masses that the 
neutrino oscillation experiments seem to call for. 
Similarly, a significant value for such couplings 
has been shown to provide respite from the tachyonic nature 
of sleptons in models with anomaly-mediated supersymmetry 
breaking~\cite{Allanach:2000gu}. It is thus of both theoretical and 
phenomenological interest to consider violations of $R$-parity.

Limiting ourselves to a renormalizable superpotential, 
the possible $R_p$-violating terms can be 
parametrized as 
\begin{eqnarray} 
W \supset \sum_{i} \kappa_i L_i H_2 +
     \sum_{i,j,k} \bigg (\l _{ijk} L_iL_j E^c_k+ 
\l ' _{ijk} L_i Q_j D^c_k+ \l '' _{ijk} U_i^cD_j^cD_k^c ,  
\bigg ) 
\label{super} 
\end{eqnarray} 
where $i,j,k$ are generation indices, $L$ ($Q$) denote the left-handed lepton 
(quark) superfields, and $E^c$, $D^c$ and $U^c$ are the right-handed 
superfields for charged leptons, down and up-type quarks respectively. 
The couplings $\l_{ijk}$ and $\l''_{ijk}$ are antisymmetric 
in the first and the last two indices, respectively. 
A conserved baryon number requires that all the $\l ''_{ijk}$
vanish identically, thereby avoiding rapid proton decay.

As with their $R_p$-conserving cousins, namely 
the  usual Yukawa couplings, these couplings are entirely 
arbitrary. Some phenomenological constraints exist, though. 
For example, the preservation of a GUT-generated 
$B-L$ asymmetry necessitates the preservation of 
at least one of the individual lepton numbers over 
cosmological time scales~\cite{Dreiner:vm}. At a more 
prosaic level, the failure of various collider experiments~\cite{nosusy} 
to find any evidence of 
supersymmetry\footnote{Although $R$-parity violation has been 
        touted as an explanation~\protect\cite{HERA_expl}
        for the reported excess of      high-$Q^2$ events at 
        {\sc hera}~\protect\cite{HERA}, it is no longer clear that 
        this anomaly persists.}
has implied constraints in the parameter space. 
Even for superpartners too heavy to be produced directly, 
strong bounds on these couplings 
can be inferred from the remarkable agreement between 
low energy observables and the SM predictions. 
These include, for example, meson decay 
widths~\cite{Barger:1989rk,Bhattacharyya:1995pq}, 
neutrino masses~\cite{neutrino_mass,Bhattacharyya:1995pq}, 
rates for neutrinoless double beta decay~\cite{bb0nu}, etc. 
The bounds generally scale with the sfermion mass and, for 
$m_{\tilde f} = 100 \gev$, they range from $\sim 0.02$ to $0.8$~\cite{rplimits}.

In general, therefore, when one studies the collider signals for 
$R_p$-violating supersymmetry, it is usual to consider the so-called 
`weak' $R_p$-violation scenario where the production of the superparticles 
goes through gauge couplings and the only role of $R_p$-violation is in the 
decay of the lightest supersymmetric particle~\cite{lspdecay}. Such
studies are clearly insensitive to the exact size of the $R_p$-violating 
coupling as long  as it is large enough to make the decay length 
of the LSP undetectable
\footnote{If any of the $R_p$-violating 
                couplings is $> 10^{-6}$ or so, then the
                LSP will decay within the 
                detector~\protect\cite{Dawson:1985vr}.}. 
The processes that are directly sensitive to the size of this 
$R_p$-violating coupling are the production of sparticles 
through them~\cite{Moreau:2000bs,HERA_expl}, the 
decays of sparticles through them~\cite{Ghosh:1997bm,hanmargo}
or through indirect effects of the virtual sparticle exchange by interference 
of the $R_p$-violating amplitude with the SM one\cite{Bhattacharyya:1994yc,
Kal, Hewett:1997ce, Ghosh:1997bm,Hikasa:1999wy}. 

An inspection of the aforementioned low-energy constraints shows that they are 
strongest when the term involves only the first two generations, 
and are often rather weak when one or more of the superfields belong 
to the third generation. In the context of collider experiments, 
a measurement of such couplings could, in principle, be done in 
more than one way, e.g., direct production, the decays of particles such as
the top quark or sparticles other than the LSP through $\rp$ coupling,
as well as through virtual exchanges. For example, in the event of a
large coupling, one could study the rate for the production of a single
superparticle. 
However, such measurements are subject to uncertainties from the
luminosity measurement in some cases and also from the knowledge of branching
ratios.  We therefore revisit the issue of the indirect determination of
some of the $\lambda'_{ijk}$ couplings at the LHC, using their contribution
to the production of lepton pairs.

Clearly, for such contributions to be significant, the initial state 
must involve quarks of the first generation or, in other words, at least 
one of $j$ and $k$ must be 1. Restricting ourselves, for the sake 
of definiteness, to muons in the final state, one sees that the  
least constrained among the relevant
couplings~\cite{rplimits} are $\l'_{231}$
and $\l'_{211}$.  We look at the reach of the LHC in the two-dimensional
$\l'$--$m_{\tilde q}$ plane. The current study goes much beyond 
the earlier analyses of this 
process~\cite{Bhattacharyya:1994yc,Hewett:1997ce} at the Tevatron in that 
we exploit the differences in
the angular distribution of the leptons  as well as the invariant mass 
distribution of the lepton pair. The use of this additional information
increases the sensitivity  to the $R_p$-violating couplings. Through a 
detailed experimental analysis, we show that it should be possible to keep 
the systematic uncertainties at the level of a few per cent. We further find 
that this analysis would serve a role complementary  to that played by looking 
for $\rp$-violating decays of the squarks.

The rest of the article is planned as follows. 
In the next section, we first outline the details of
the production of $\mu^+ \mu^-$ pairs at the LHC along with the  
discriminatory features of the various distributions we  use. We present the
result of a parton-level calculation in this discussion. Following it up 
are the details of the maximum likelihood analysis technique used by us, and we
present the results of the same at the parton level as well as at the level of
fully generated events. We then give a discussion of the various systematic 
uncertainties in the analysis. After this, we briefly discuss the 
$R_p$-violating branching ratios of the third-generation squark. We show that 
for the squark mass range considered in this paper,  there exists
a region in parameter space in which the $\l'_{231}$ 
coupling could in principle be measured at the $5\%$ level at LHC at full 
luminosity.  This demonstrates the above mentioned complementarity very nicely.

\section{Drell--Yan in the presence of $R_p$ violation}
The Drell--Yan process at a hadronic collider has, over the 
years, been studied in great detail. Its analysis has served 
as an excellent theoretical laboratory, first for the quark 
parton model and later for perturbative QCD. In fact, not only the NLO, but 
even the NNLO corrections have been computed~\cite{nnlo}. 
In the context of the present analysis, though, we shall
limit ourselves to leading-order 
calculations. The reasons are manifold. For one, the NNLO calculations,
within the SM, have demonstrated that a constant $K$-factor gives a 
very good description of the corrections over a very wide range of 
kinematic variables. Secondly, the radiative 
corrections in the presence of non-SM physics, such as the case under
study, have not yet been calculated. It might be argued, 
though, that since we are interested in probing typically
small $\rp$ 
couplings, a very precise calculation of the strong corrections 
to these additional contributions is not crucial. Instead, we 
might assume that the $K$-factor is essentially the same 
as in the SM. Clearly, this estimate is quite a reasonable 
one; it is supported by the recent calculations~\cite{Beenakker:1999xh}
of $K$-factors for chargino or neutralino pair-production, each of which  
receives $t$-channel contributions as in the present case. 
This assumption simplifies the analysis, as the Born term is extremely simple 
to analyse.

A look at the superpotential given by Eq.(\ref{super}) tells us that 
it is only the $\l'_{ijk}$ couplings that can affect Drell--Yan
production of a dilepton pair.  Expressed in terms of the 
component fields, the relevant part of the  Lagrangian reads
\be
\barr{rcl}
{\cal L}_{\lambda'} & = & \dis 
  \lambda'_{ijk} \bigg[ \,
   {\tilde \nu}^i_L {\bar d}^k_R d^j_L
 + {\tilde d}^j_L {\bar d}^k_R \nu^i_L
 + ({\tilde d}^k_R)^\ast ({\bar \nu}^i_L)^c d^j_L   
         \\ 
&& \hspace*{2em} \dis 
        - {\tilde e}^i_L {\bar d}^k_R u^j_L
   - {\tilde u}^j_L {\bar d}^k_R e^i_L
   - ({\tilde d}^k_R)^\ast ({\bar e}^i_L)^c u^j_L \,\bigg]
 + {\rm h.c.} \ .
\earr
\ee
A non-zero $\l'_{2jk}$ would then lead to an additional
$u$-channel ($\tilde d^k_R$ exchange) diagram for the 
process $\bar u_j u_j \rightarrow \mu^- \mu^+$ and 
a $t$-channel ($\tilde u^j_L$ exchange) diagram for 
$\bar d_k d_k \rightarrow \mu^- \mu^+$. Note that 
neither resonance production processes leading to the dimuon
final state nor processes such as $q_i \bar q_j \rightarrow e^- \mu^+$
can occur when only a single $\rp$ coupling
is non-zero, an assumption~\cite{fcnc} that we shall work 
with. 

The differential cross section is modified in a straightforward way and reads, 
in the centre of mass  of the $\mu^+ \mu^-$ system, as
\be
\barr{rcl}
\dis \frac{\mbox{d}\hat{\sigma}}{\mbox{d}\cos\theta}[q\bar{q}
\rightarrow \mu^- \mu^+]
 & = & \dis \frac{\pi\alpha^2\hat{s}}{24} \Bigl\{
        (1+\cos\theta)^2
        \left[ |f^s_{LR}|^2 + |f^s_{RL}|^2 \right] 
         \\[1.5ex]
& & \dis \hspace*{4em} 
        + (1-\cos\theta)^2\left[|f^s_{LL}|^2+|f^s_{RR}|^2\right]
 \Bigr\}
\\[1.5ex]
f^s_{LR} & = &\dis - \: \frac{Q^q}{\hat{s}} + \frac{g_L^q
  g_L^e}{\hat{s}-m^2_Z+i\Gamma_Zm_Z}
     \\[1.9ex]
f^s_{RL} &=& \dis - \: \frac{Q^q}{\hat{s}}
             + \frac{g_R^q g_R^e}{\hat{s}-m^2_Z+i\Gamma_Zm_Z}
     \\[1.9ex]
f^s_{LL} &=& \dis -\frac{Q^q}{\hat{s}}
             + \frac{g_L^qg_R^e}{\hat{s}-m^2_Z+i\Gamma_Zm_Z}
-\frac{1}{2}\, 
        \frac{(\lambda'_{2jk}/e)^2}{\hat{u}-m^2_{\tilde{d}_{kR}}}
\delta_{q u_j}   
        \\[2.9ex]
f^s_{RR} &=& - \dis \frac{Q^q}{\hat{s}}
             +\frac{g_R^q g_L^e}{\hat{s}-m^2_Z+i\Gamma_Zm_Z}
+\frac{1}{2}\, 
        \frac{(\lambda'_{2jk}/e)^2}{\hat{t}-m^2_{\tilde{u}_{jL}}}
\delta_{qd_k} \ ,
\earr
\label{eq:ideal}
\ee
where $Q^f$ represents the charge of the fermion $f$ and $g^f_{L, R}$ 
its couplings to the $Z$. 
Clearly, the new contribution is relevant only if 
$q = u, d$ or, in other words, only if at least one of $j$ 
and $k$ refers to the first generation. Of the nine $\rp$ couplings
that could, in principle, contribute to dimuon production, we thus 
need to concern ourselves with only five. For a given strength 
of the coupling, the change wrought in the total cross section 
would depend on the mass of the squark(s) involved as well 
as on the particular subprocess that is being affected. 

%
\FIGURE[!h!]
{ \vspace*{-4ex}
\epsfig{file=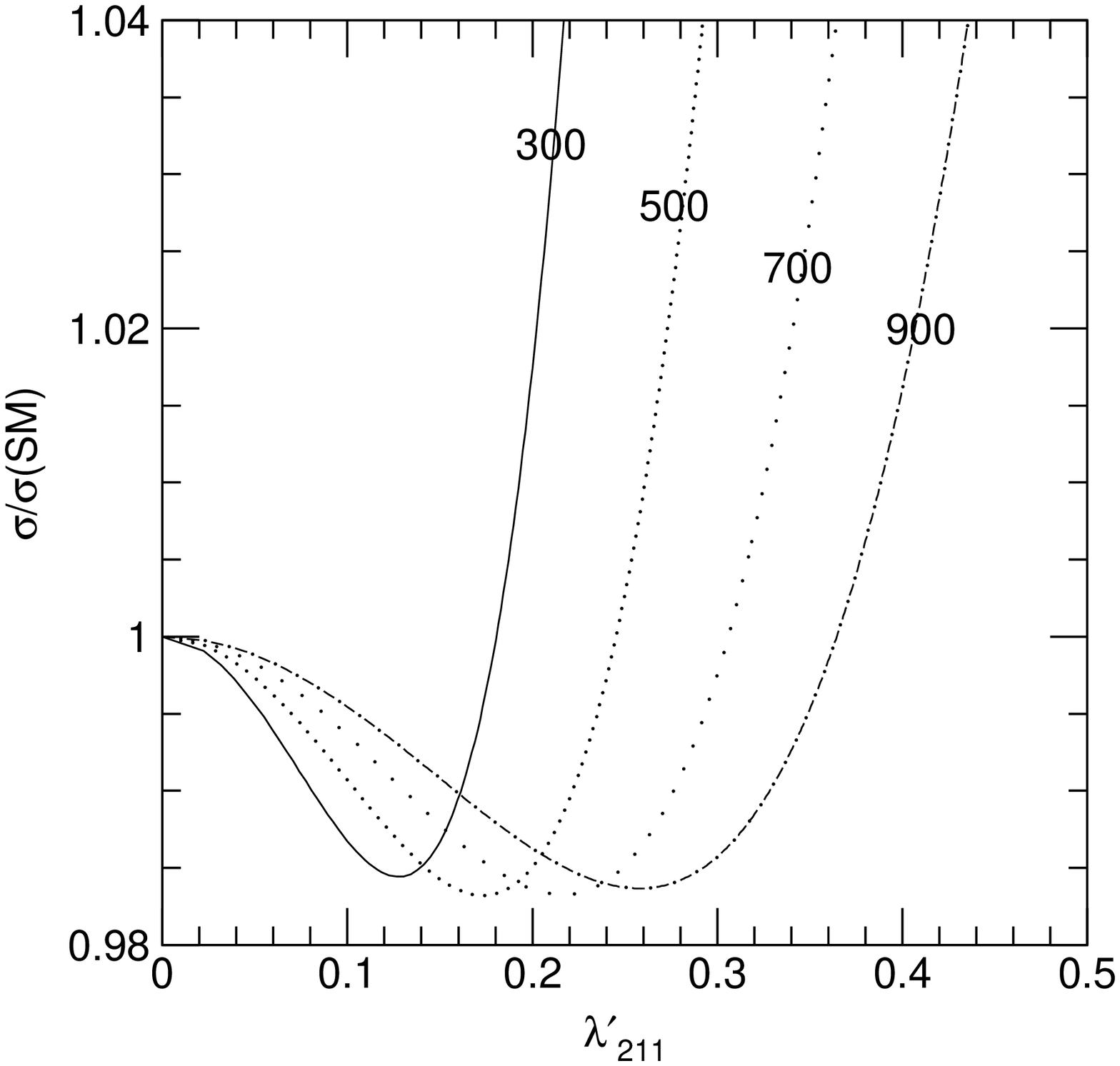,width=0.60\textwidth}
\vspace*{-2ex}
\caption{\em Relative change of the dimuon cross section 
        at the LHC as function of the coupling $\l'_{211}$.
        The assumed lower limit on the dilepton invariant
        mass is 500~GeV.
        The curves correspond to four different squark 
        masses: 300, 500, 700 and 900 ~GeV respectively.}
\label{Fig:sigfour}
}
%
In Figure~\ref{Fig:sigfour}, we present the relative deviation 
of the total cross section for a particular coupling,
i.e. $\l'_{211}$, and different values of the squark 
masses (for convenience, we assume 
that $\tilde u_L$ and $\tilde d_R$ are degenerate).
The lower cut on the dilepton invariant mass assumed
for the calculation is 500~GeV.
As far as the total cross section is concerned, the difference 
is only at the level of a few per cent, even for moderately 
large values of the coupling. However, this is somewhat misleading, since 
the deviation from the Standard Model increases with the dilepton mass, and 
therefore the observed integrated effect strongly depends on the 
lower limit on the dilepton mass taken for the calculation. 
The extra contributions manifest themselves more markedly 
in the kinematic distributions, modifying them from those 
expected within the SM in three essential ways:
(1) enhanced cross section for high $\ell^+\ell^-$ 
invariant mass, (2) different lepton angular distribution in the 
$\ell^+\ell^-$ rest frame and (3) different boost distribution of the 
$\ell^+\ell^-$ system.

\FIGURE[htb]{
\centerline{
\epsfxsize=5.6cm\epsfysize=5.6cm
                     \epsfbox{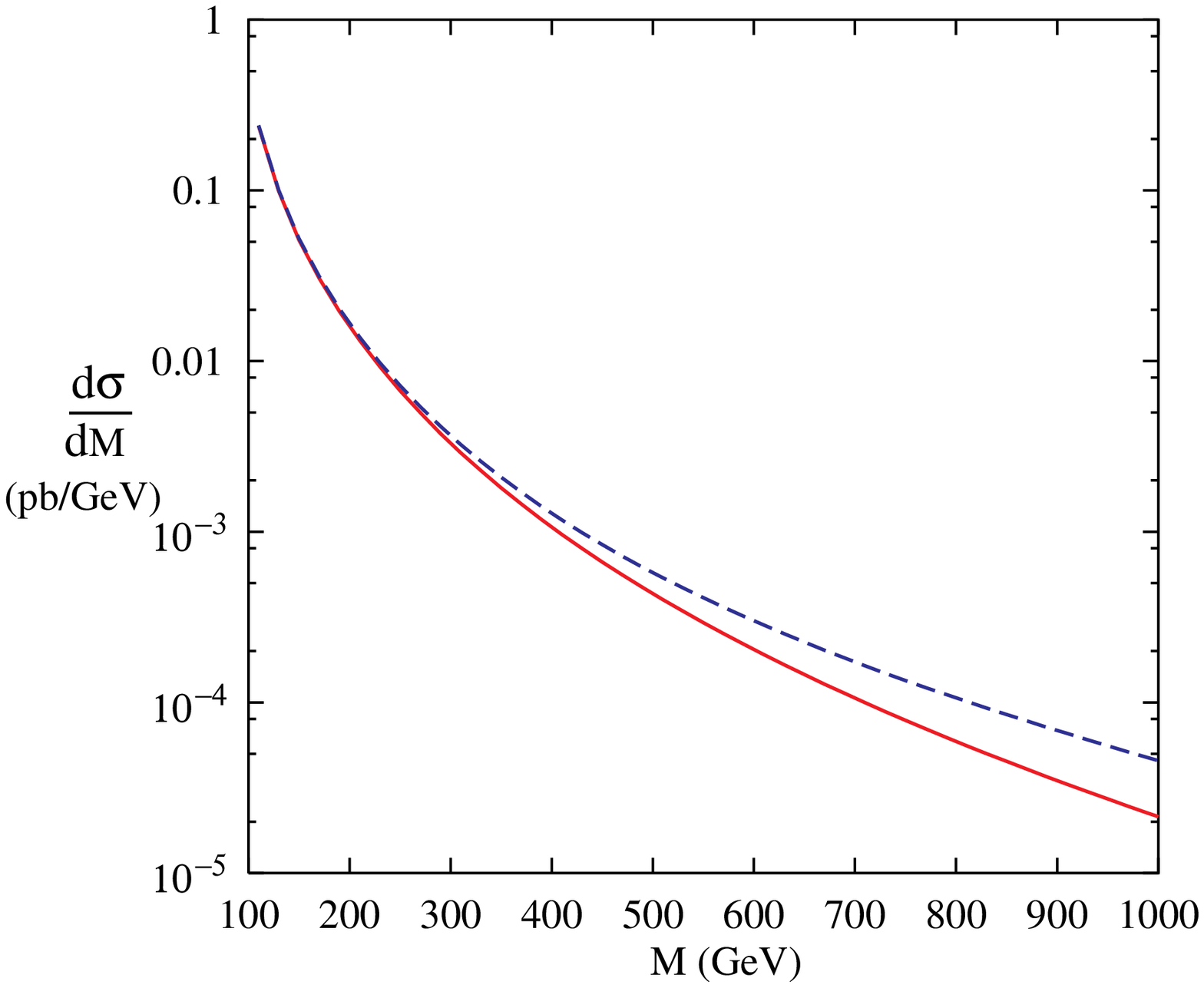}
\hspace*{-0.8cm}
\epsfxsize=5.6cm\epsfysize=5.6cm
                     \epsfbox{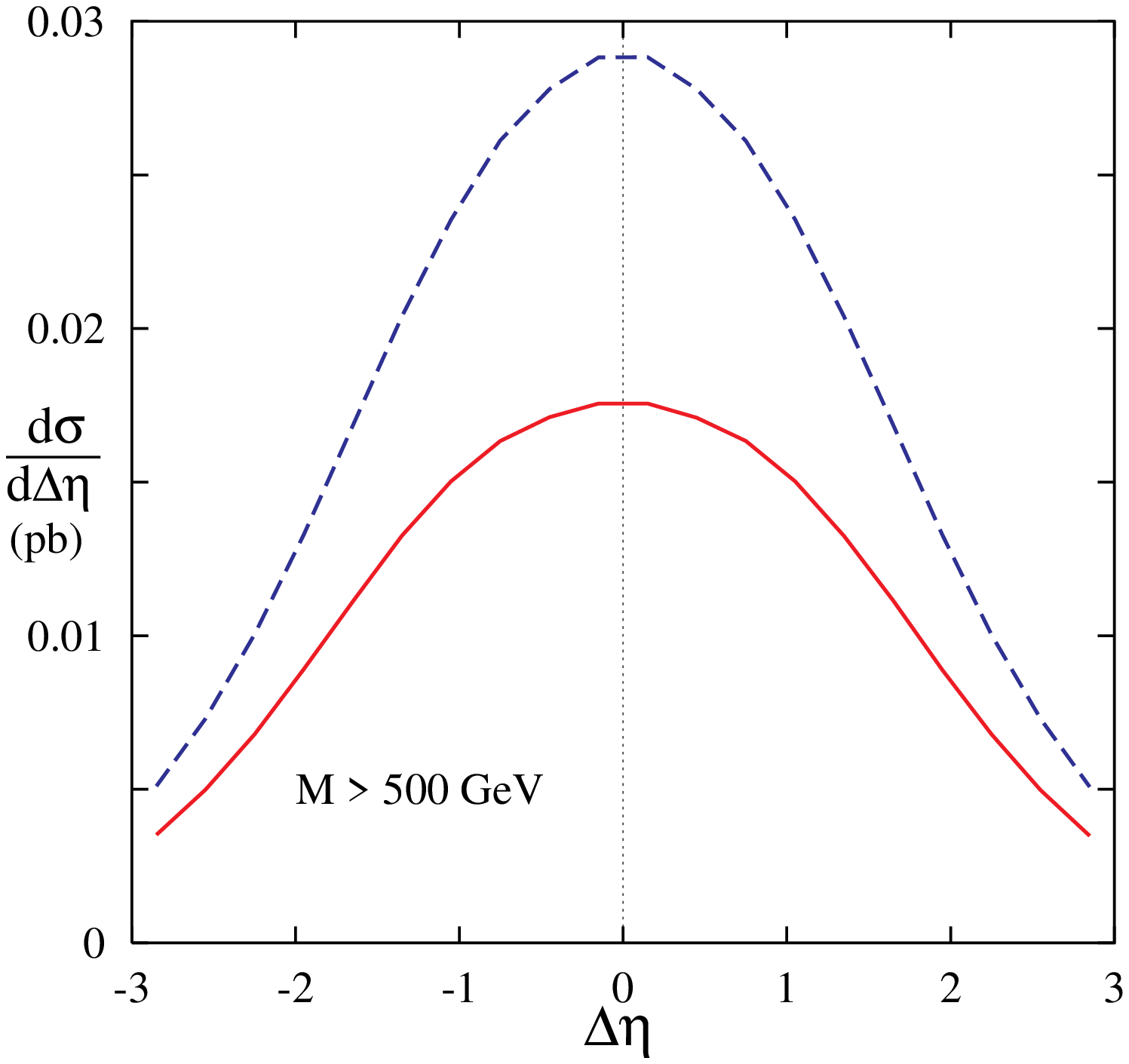}  
\hspace*{-0.8cm}
\epsfxsize=5.6cm\epsfysize=5.6cm
                     \epsfbox{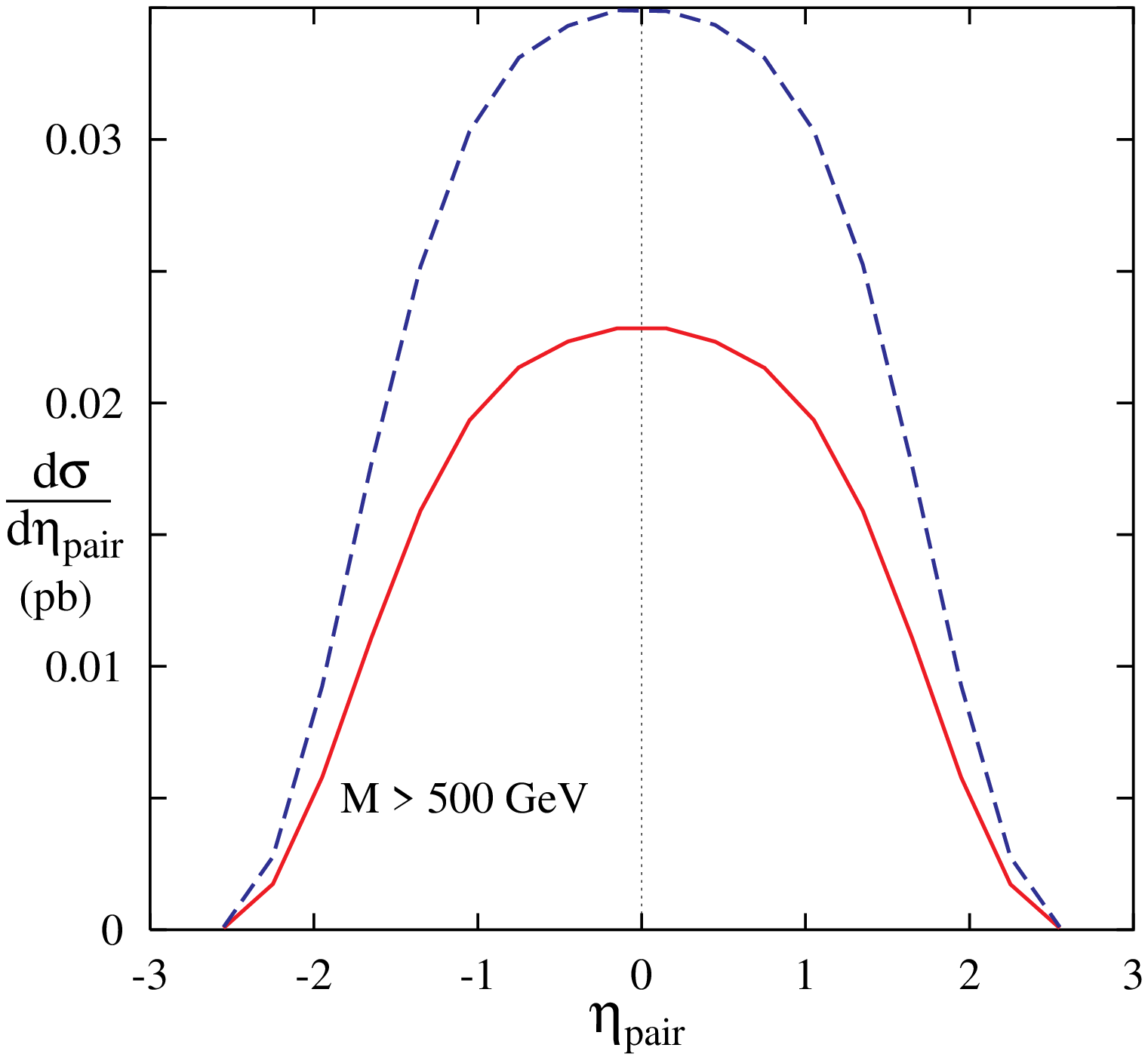}
}
\caption{\em The phase-space distributions for the Drell--Yan process 
        at the LHC. We have demanded that the $\mu^\pm$  have a 
        minimum transverse momentum of 50 GeV each and be within the 
        pseudorapidity range $-3 < \eta < 3$.
        For the rapidity distributions, an additional cut ($M > 500 \gev$) 
        has been imposed. In each case, the lower curve corresponds to the 
        SM and the upper curve to 
        $\lambda'_{211} = 0.5$, $m_{\tilde q} = 800 \gev$.
        }
    \label{Fig:distribs}
}

For a given point in the supersymmetric parameter space, these differences 
are demonstrated in Figure \ref{Fig:distribs}. The figure shows distributions 
in the invariant mass $M$ of the dimuon pair, the difference in the 
rapidities of the two leptons $\Delta \eta$ and the rapidity of the
pair itself $\eta_{pair}$. Clearly, the use of the 
differential distributions would result in an enhanced sensitivity 
with respect to just the event rate comparison of Figure \ref{Fig:sigfour}. 
Note the substantial broadening of the distribution in $M$. 
In fact, rather than limiting ourselves to a study of a 
distribution in one of the three independent variables listed above, 
it is conceivable that the use of the full kinematic information,
taking into account the correlation of the kinematic variables,
would be even more powerful.  In the next two sections we discuss how exactly 
we propose to do this, albeit with a choice of independent variables slightly 
different from that mentioned above.

\section{Maximum likelihood method}

Our goal, then, is to exploit to the full the differences 
in the kinematic distribution of the $\mu^+ \mu^-$ pair caused by the 
presence of  the $\rp$ interactions. Such a study is expected 
to yield a measurement (or at least constraints) in the 2-dimensional 
supersymmetric parameter space, namely the mass-coupling plane. 
The problem of simultaneous determination of these two, viz. $\l'$ and 
$m_{\tilde q}$, is the
classical problem of parameter estimation. In view of the systematic 
uncertainties (experimental due to the luminosity and theoretical due to 
parton distributions, $K$-factors, etc.) in the normalisation of the signal,
it is best to use the maximum likelihood method, which does not require a
precise knowledge of the absolute size of the signal, but only that of its
shape. 

As Figure \ref{Fig:distribs} shows, the fall-off in 
the invariant mass distribution is indeed  substantially slower
for the $R_p$-violating (RPV) contribution. For the other distributions too, 
the 
difference is quite discernible. However, instead of using the variables 
of Figure \ref{Fig:distribs}, we choose to work with an equivalent 
(independent) set, namely the momentum fractions $x_1$ and $x_2$ of the 
initial-state partons and the cosine of the scattering angle, $\cos \theta$. 
Neglecting the transverse momentum of the $\mu^+\mu^-$ system, the event
kinematics for a given event $i$ is completely specified  by these three
variables. The log likelihood function
is defined as:
\begin{equation}
\ln{\cal L}=\sum_{i=1}^{N_{\rm ev}}
        {\ln F(\lambda^\prime,m_{\tilde q}, x_1^i,x_2^i, \cos\theta^i
)} \ ,
\label{eq:like}
\end{equation}
with $F$  given by:
\begin{equation}
F=\frac{1}{\sigma(\lambda^\prime,m_{\tilde q})}
        \;
\frac{{\rm d} \sigma}{{\rm d} x_1 {\rm d} x_2 {\rm d} \cos\theta}
        (\lambda^\prime,m_{\tilde q}, x_1^i,x_2^i, \cos\theta^i) \ .
\label{eq:weight}
\end{equation}
The cross section $\sigma(\lambda^\prime,m_{\tilde q})$ in the denominator is
the one  obtained {\em after} imposing all the analysis cuts.  The latter,
of course, are chosen so as  to maximize the RPV contribution in the signal.   
The best estimate of the true values of the two parameters is then given by the
particular pair ($\lambda^\prime, m_{\tilde q}$) that maximizes 
$\log{\cal L}$ 
for a given data sample. In addition to the advantage that comes from using
only the shape of distributions as mentioned  above, this method 
also has the good features of not requiring any binning, as well as 
exploiting the
correlations among the different variables optimally. We note here that the 
objective of avoiding the systematic (both theoretical and experimental) errors
due to the imprecision in the knowledge of the absolute size of the signal,
could, in principle,
also be achieved by comparing the size of the dimuon  signal with the 
dielectron one. However, the latter method presupposes 
that no unknown physics effects exist in the dielectron mass spectrum and
hence it is not completely model-independent. Furthermore, the comparison of
two different spectra involves the compounding of errors, thereby 
affecting the accuracy adversely.

\section{Data Analysis}
For our analysis of the simulated data, 
we first generate events corresponding to an integrated luminosity of
$100~{\rm fb}^{-1}$, which corresponds to a year of LHC operation in its
high luminosity mode.  The evaluation of the achievable sensitivity in the
$\lambda^\prime$ measurement requires the generation of many times the 
statistics expected for a single experiment, for a fine grid on 
$\lambda^\prime$ for a few values of $m_{\tilde q}$. The parton-level 
generation of unit weight events without initial-state QCD showering is about
ten times faster than the generation of full PYTHIA  \cite{PYTHIA}
events followed by the fast detector simulation for
the ATLAS detector \cite{ATLFAST}. We will therefore first
evaluate the statistical sensitivity of the experiment at the parton level. 
The results thus obtained will then be compared with the same procedure applied
to fully generated events for a few selected points in parameter space,
in order to evaluate the effect on  the experimental sensitivity of the 
experimental smearing and of the introduction of initial-state QCD radiation.

We  select events  with  two isolated opposite-sign muons, 
satisfying the following two requirements: 
1) $P_T^{\mu}>10$~GeV,   $|\eta_{\mu}|<2.5$ and 2) $m_{\mu^+\mu^-}>500$ GeV.
The first criterion essentially  ensures that the
muons are visible in the detector. The invariant-mass
cut, on the other hand, is motivated by the observation
that the relative deviation of the cross section
starts to become significant only at
$m_{\mu^+\mu^-} \sim m_{\tilde q}$~\cite{Bhattacharyya:1994yc}.
The optimal choice of the cut is determined not only
by the squark mass under consideration, but also by the total number of events
(in other words, the luminosity). However, rather than
working with a squark mass-dependent cut, we choose to adopt the
simpler strategy of a fixed choice for the invariant-mass cut.
Approximately 7500 events survive these cuts  for our choice of 
the luminosity.

As is clear from the definition of the likelihood function, a complete 
knowledge of the kinematics  of the event is necessary for its calculation.  
However, it must be borne in mind that, at a $pp$ collider such as the
LHC, it is not possible to know for certain the initial direction
of the quark (in the $\bar qq$ hard scattering subprocess).
Hence, only the absolute value of $\cos\theta$ is measurable, not the sign.
Part of this information can, however, be recovered, by using
the knowledge of $x_1$ and $x_2$ and the fact that, in the
proton, the $x$ distribution for valence quarks is harder
than for antiquarks.  Arbitrarily labelling the proton beams 
``1'' and ``2'', the difference $x_1 - x_2$ can be inferred from 
the longitudinal momentum of the dimuon pair.
\FIGURE[!h]{
\centerline{
\dofigs{0.45\textwidth}{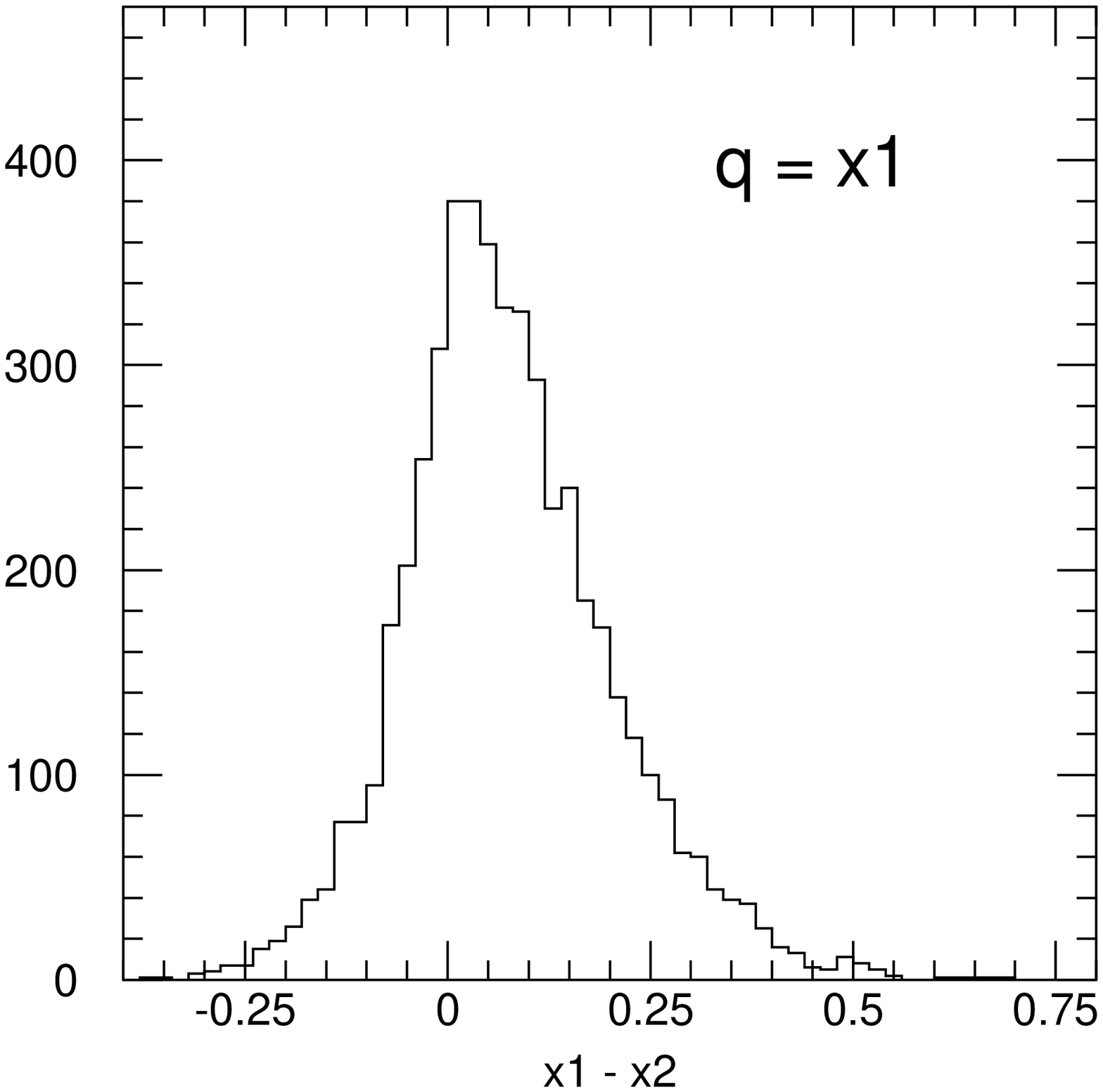}{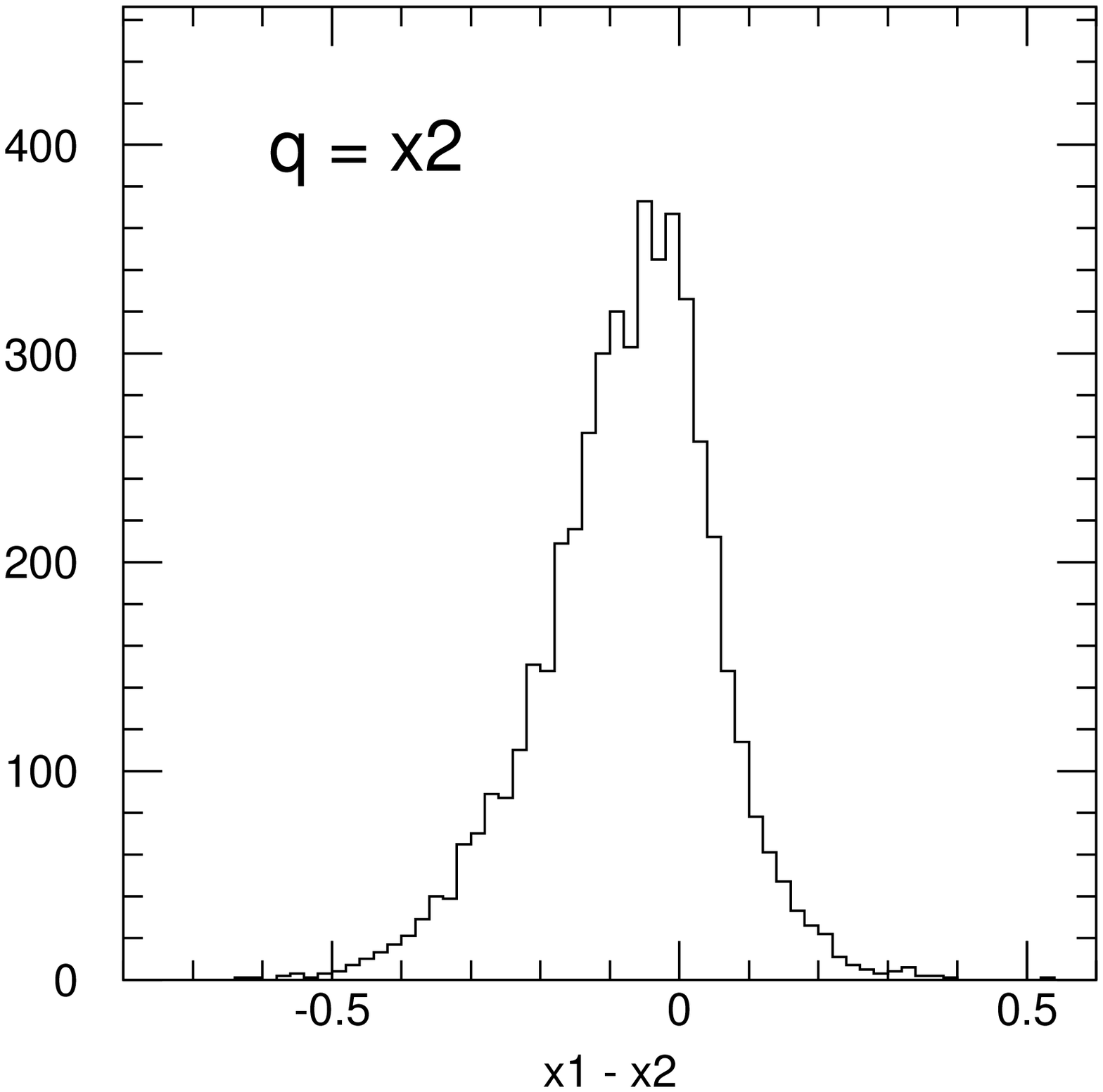}}
\caption{\em Difference $x_1-x_2$ when the quark is picked up from
the proton labelled ``1" (left plot) and ``2" (right plot).}
\label{Fig:signth}
}

In Figure \ref{Fig:signth} we show the
distribution of $x_1-x_2$, when the quark is picked up from
the proton labelled ``1" (left plot) and ``2" (right plot). It can be
clearly seen that in most cases the quark is taken from the side with higher 
$x$.  To make use of this information,
the likelihood is built by summing the functions $F$
calculated for both signs of $\cos\theta$, weighted by the probability,
for the given ($x_1$, $x_2$) combination that the quark is in proton
``1" or ``2".

\subsection{Parton level analysis}
For the sensitivity study, we generated a grid of points
with different values of $\lambda^\prime$ (with a spacing of $0.025$)
for four different equispaced values of the squark mass ranging from 
300 to 900~GeV. 
For each point, $5 \times 10^5$ events with unit weight were generated.
The likelihood was calculated in steps of $\l^{\prime 2}$ for 
$-0.05 \le \lambda^{\prime2} \le 0.2$. The inclusion of the unphysical range
is necessary for the evaluation of the confidence interval.
If the likelihood function has a local maximum for a positive (physical) 
$\l^{\prime 2}$, we take this value as {\em a measurement}. Otherwise, 
we take the absolute maximum, even though it may be 
in the unphysical (negative) 
region. The cross section being a quadratic function of 
$\l^{\prime2}$, a secondary maximum may appear, 
and often does for negative $\l^{\prime 2}$. 
In fact, even for $\l^{\prime 2}$ values large 
enough for the experiment to be sensitive to squark exchange, 
a small fraction  of the Monte Carlo experiments can show the unphysical 
maximum to be higher than the physical one. 
This feature is  illustrated in Figure \ref{Fig:pllik2}, 
where we display the likelihood as a function of 
$\lambda'^2_{211}$ for four experiments, assuming 
$m_{\tilde q}=500$~GeV.  The upper plots are for $\l'_{211}=0.15$, which is 
below the experimental sensitivity for this particular 
squark mass.
In this case, even when the physical maximum is 
found, the dip between the two maxima is shallow 
(\mbox{1$\sigma\sim\Delta\ln {\cal L}=0.5$}), and 
very often the absolute
maximum occurs at a  negative $\l^{\prime 2}$.
The lower plots are for  $\l'_{211}=0.2$. The two 
maxima here are always well separated; the rare 
cases when the absolute
maximum is in the unphysical region correspond to 
experiments for which the 
positive maximum is somewhat displaced from the nominal value.
\vspace*{-2ex}
\def\bottomfraction{1.}
\def\textfraction{0.}
\FIGURE[!h]{
\dofig{0.9\textwidth}{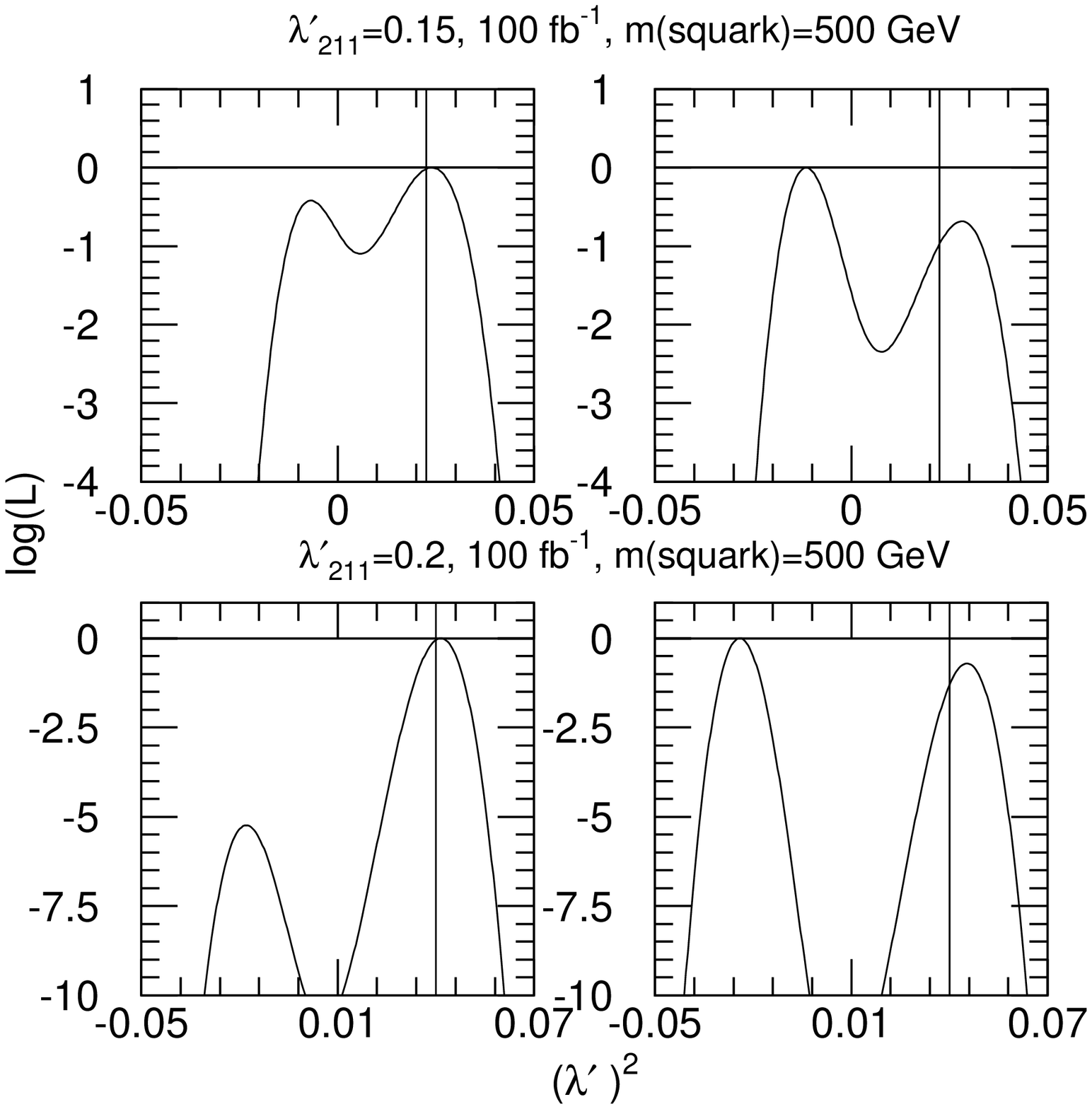}
\vspace{-1cm}
\caption{\em Shape of the log likelihood function as a function
of $\lambda'^2_{211}$ for four different Monte Carlo experiments.
The value of $\lambda'_{211}$ used in the generation of events, shown as a 
vertical line, is 0.15 for the upper  
plots, and 0.2 for the lower plots. 
The assumed squark mass is $m_{\tilde q}=500$~GeV.
The integrated luminosity is 100~fb$^{-1}$.}
\label{Fig:pllik2}
}
%

In order to evaluate the uncertainty on the $\lambda^\prime$ measurement, 
we generated $\sim 1500$ Monte Carlo experiments for an integrated 
luminosity of 100~fb$^{-1}$; for each of them we calculated the
maximum of the likelihood function according to the 
above prescription. Since the generated statistics is 
only $\sim 70$ times the statistics for a single experiment, each of the 
Monte Carlo experiments was produced by randomly picking, inside the available 
statistics, the $\sim 7500$ events corresponding to one year of running.
With this procedure each event is used for $\sim 25$ Monte Carlo 
experiments.
%
\def\topfraction{1.}
\def\textfraction{0.}
\FIGURE[!h]{
\centerline{
\dofig{\textwidth}{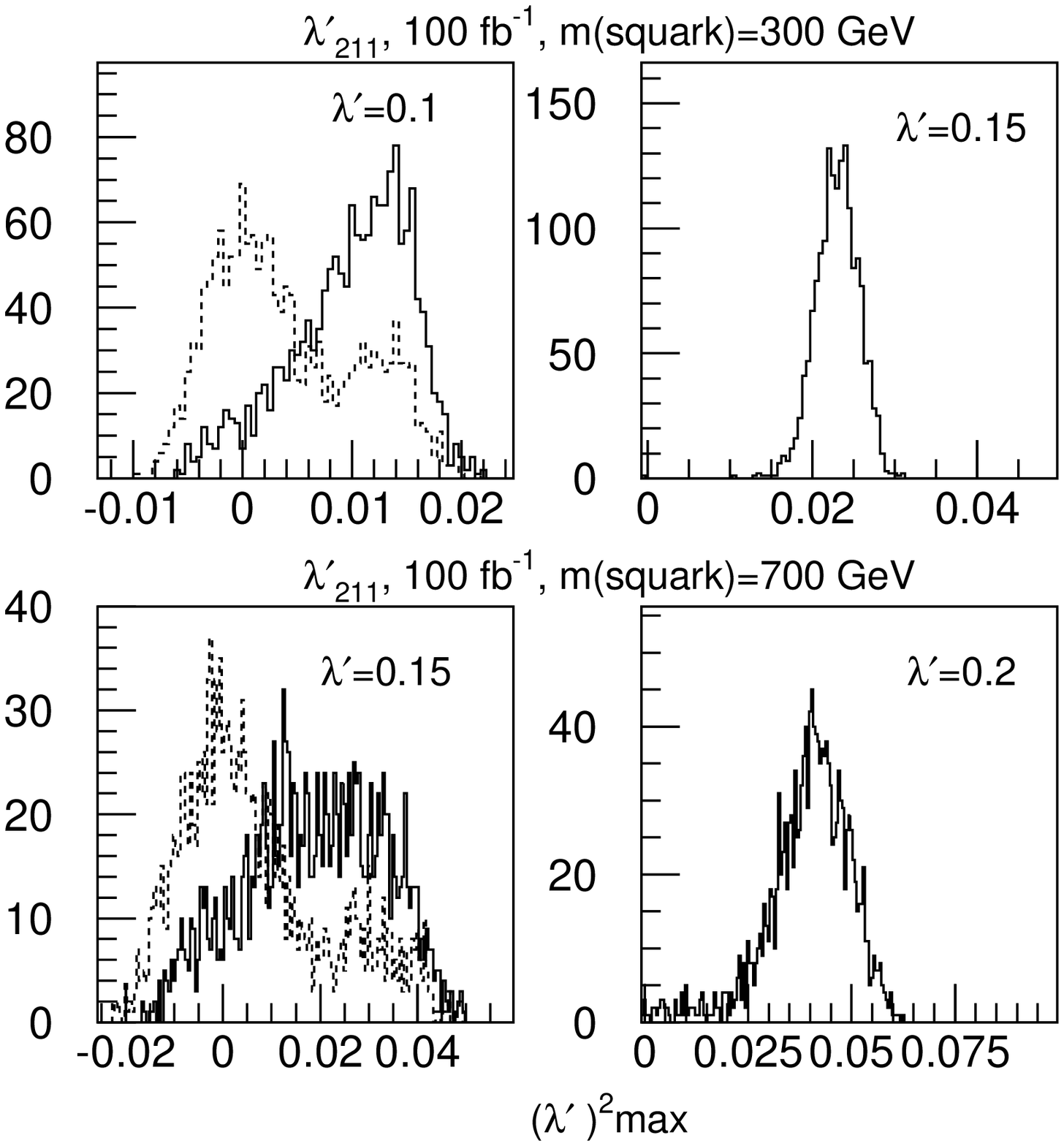}}
\vspace{-1.4cm}
\caption{\em Distribution of the $\lambda'^2_{211}$ value 
estimated through the maximization of the likelihood function 
for a set of $\sim$1500 Monte Carlo experiments. 
The upper (lower) set corresponds to a squark mass 
of 300 (700) GeV.  For each case, we show the distributions for three 
values of $\lambda^\prime$ used in the generation of events: 0 (dashed line), 
a value below the experimental sensitivity for the assumed squark mass and 
integrated luminosity (left plot, full line),
and a value for which a good $\l$
measurement is achievable (right plot).
The integrated luminosity is 100~fb$^{-1}$. }
\label{Fig:pl300700}
}
%

A common behaviour is observed for all four squark masses considered. 
For values of $\l^{\prime 2}$ approximately up to the minimum 
in the cross section shown in  Figure~\ref{Fig:sigfour}, 
the distribution of the estimated values of $\l^{\prime 2}$ 
is very broad, and extends up to a specific value of $\l^{\prime 2}$,
which depends on the mass, in such a way that the distribution of the measured 
$\l^{\prime 2}$ becomes gaussian only for $\l^{\prime 2}$ values above 
this. This means that the experiments start being sensitive to the effect 
somewhere in this transition region.
Figure \ref{Fig:pl300700} shows the results for $m_{\tilde q} = 300$ and $700$ 
GeV.

\subsection{Evaluation of experimental sensitivity}

We follow here the frequentist approach of \cite{FC}, where 
the sensitivity of an experiment is defined as:
``the average upper limit that would be obtained by an ensemble of experiments
with the expected background and no true signal".
For each considered value of the squark mass, we build, therefore, 
confidence belt according to the Neyman construction. 
We adopt as the auxiliary choice for the definition of the 
belt the one leading to ``upper confidence intervals",
defined by Eq. (2.5) in \cite{FC}. 
For an ensemble of experiments with $\lambda=0$, we then calculate the 
respective upper limits using the confidence belt thus built, and 
then take the average.
The results are shown  on the 
($m_{\tilde q}$--$\lambda^\prime$) plane in Figure \ref{Fig:plres}, 
together with the region corresponding to the bounds on $\lambda'_{211}$ from 
low-energy processes.
%
\FIGURE[!h]{
\centerline{
\dofig{0.7\textwidth}{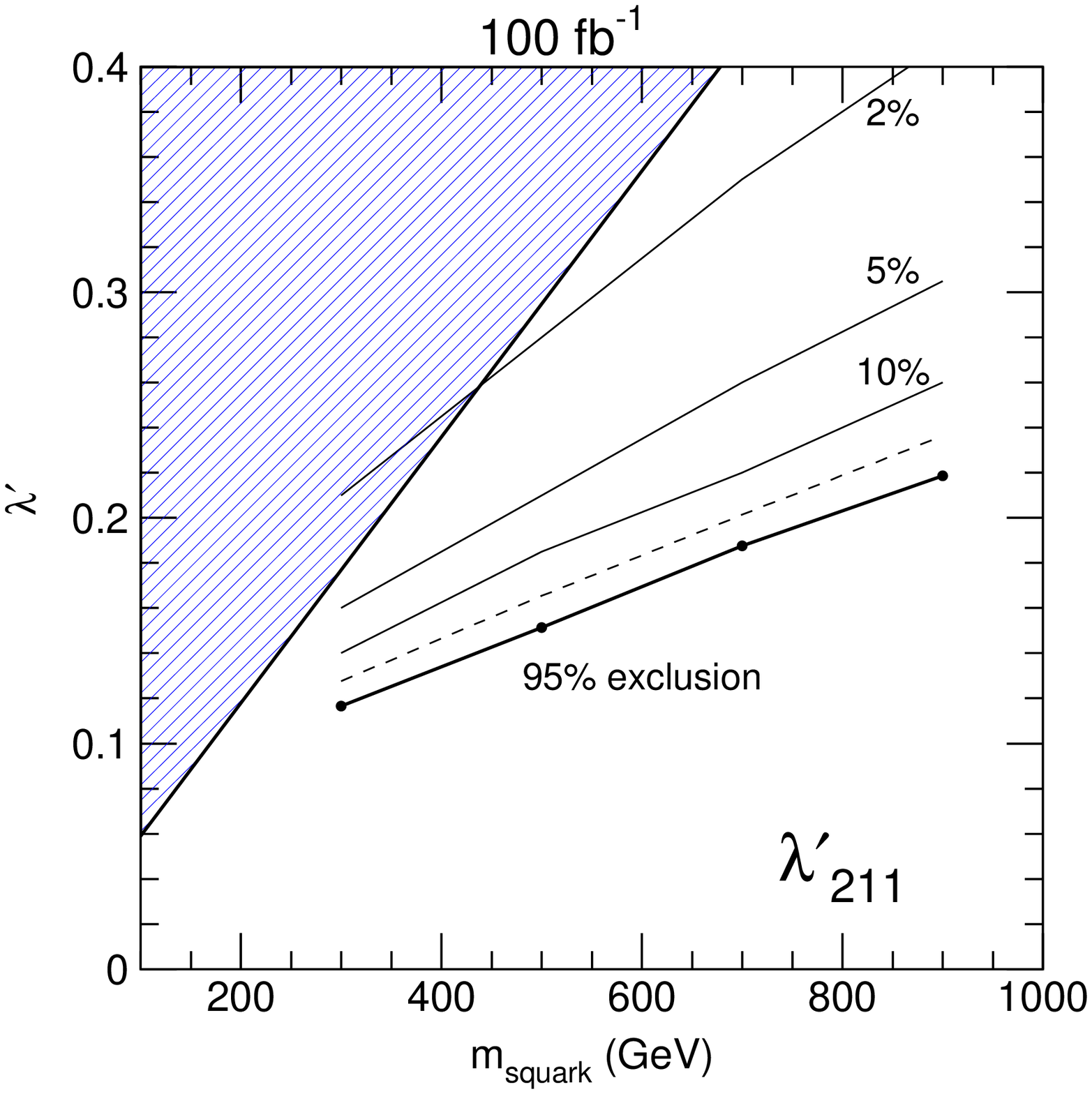}}
\vspace*{-0.8cm}
\caption{\em 95\% sensitivity region in  
the $m_{\tilde q}$--$\lambda^\prime$ plane for an integrated luminosity 
of 100~pb$^{-1}$. The dashed line corresponds to the choice of the
central interval in the Neyman construction. The analysis 
is performed at the parton level. The contours corresponding to an 
uncertainty on $\lambda^\prime$ of 2, 5 and 10\%  
(under the assumption of perfectly known squark mass) 
are also shown. The shaded region on the left is excluded
by low energy measurements.
}
\label{Fig:plres}
}
%
If no signal is present in the data, the result of the experiment
will be the exclusion of the region in the ($m_{\tilde q}$--$\lambda^\prime$) 
plane above the curve labelled `95\% exclusion'. 
If a signal is present, the experiments will be able to extract
a measurement of the $\lambda^\prime$ couplings.  
The assessment of the precision 
of this measurement depends on the available information on $m_{\tilde q}$, 
and on the dependence of the measured value of $\lambda^\prime$
on the assumed squark mass.
As a first, unrealistic, approximation we show in Figure \ref{Fig:plres}
the curves in the  ($m_{\tilde q}$--$\lambda^\prime$)  plane corresponding to 
a statistical uncertainty on $\lambda^\prime$ of 2, 5 and 10\% respectively, if
no error is assumed on the squark mass.

\vspace{-0.4cm}
\FIGURE[!h]{
\centerline{
\dofig{0.9\textwidth}{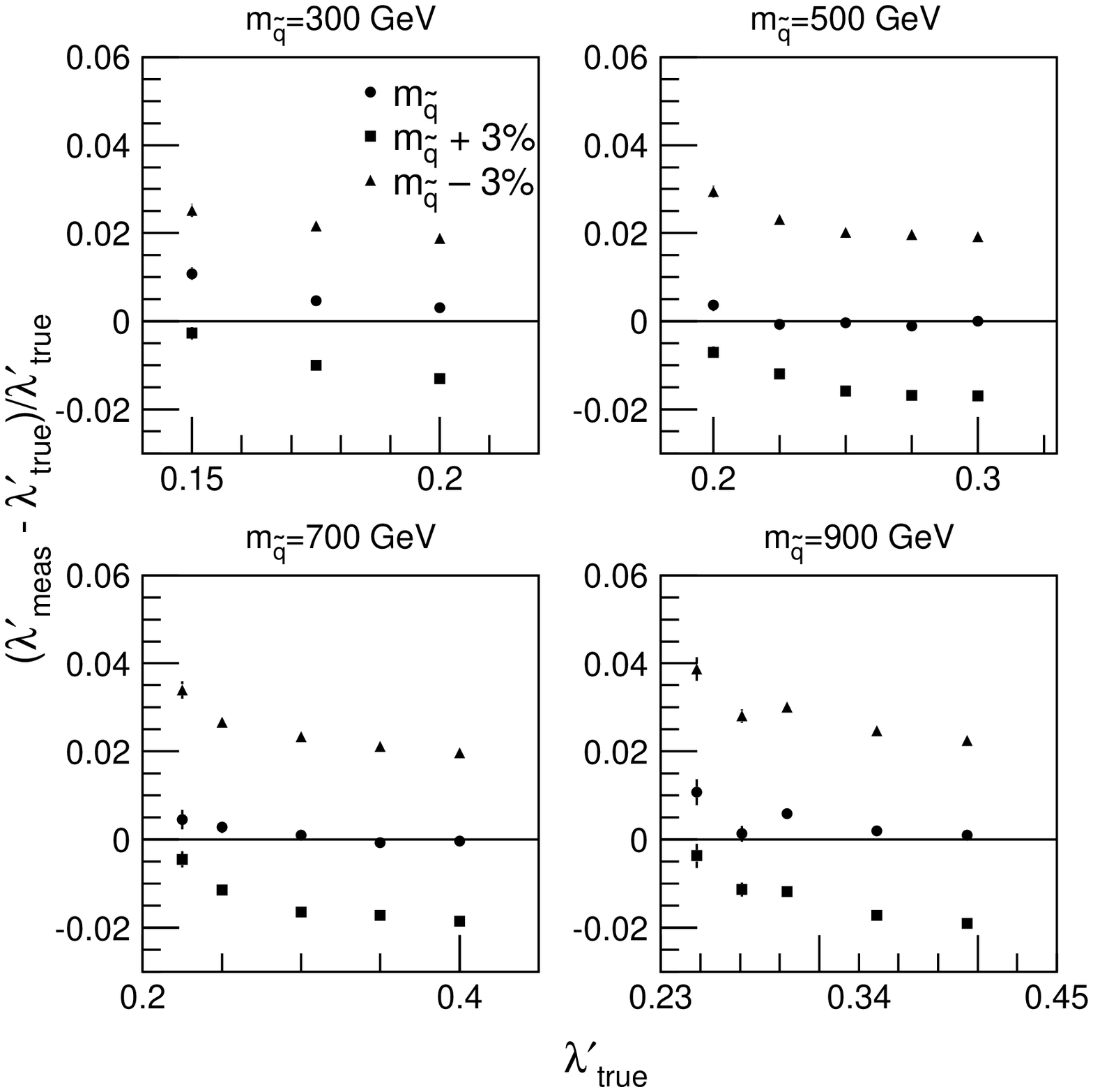}}
\vspace{-1.3cm}
\caption{\em Fractional deviation of the measured $\lambda^\prime$ 
value from the true one as a function of $\lambda^\prime$ for three 
values of the squark mass used in the likelihood 
calculation: the nominal one (dots),
reduced by 3\% (squares), augmented by 3\% (triangles).
The four plots are respectively  for the four squark masses
considered in the analysis, $m_{\tilde q}=300,500,700$ and 900~GeV}
        \label{Fig:plm}
}
It is interesting to consider the dependence of the error estimate 
on the available prior information on the squark mass. Consider 
the pair-production of the said squark at the LHC.
For the range of values of the $\lambda^\prime$  
couplings addressed in this paper, 
every supersymmetric event will contain two $\tilde\chi^0_1$ decaying
into two jets and a muon or a muon neutrino. Detailed studies have 
shown that, in the ATLAS experiment, it will be possible to reconstruct 
the  $\tilde\chi^0_1$ peak from its decay products, and, going up
the decay chain, to reconstruct the squark masses \cite{TDR}. The statistical 
precision of the mass measurement is essentially a function of 
the squark mass, and for squark masses below $\sim1$~TeV, the 
error on the mass will be dominated by the systematic uncertainty 
on the reconstruction of the parton 
energy from the jet energy measurement. Such uncertainty is estimated to be 
approximately 3\% in ATLAS \cite{parker}. Therefore, in the sensitivity region 
we have performed the $\lambda^\prime$ measurement for values of the squark
mass displaced upwards and downwards by 3\% with respect to the nominal mass. 
The results are shown in Figure~\ref{Fig:plm} for the four 
mass values considered in the analysis. The correlation between the 
$\lambda^\prime$ and $m_{\tilde q}$ measurement is positive, therefore 
a higher value of $m_{\tilde q}$ in input yields a higher $\lambda^\prime$ 
measurement. For the assumed value of the uncertainty on $m_{\tilde q}$,
the additional uncertainty on $\lambda^\prime$ from this effect is of 
order $2$--$3\%$ for all considered mass values.
One can also observe from Figure~\ref{Fig:plm} that when the nominal 
mass value is used, the bias on $\lambda^\prime$  from the likelihood fit 
is less than 1\%.

\vspace{-0.5cm}
\FIGURE[!h]{
\centerline{
\dofig{0.7\textwidth}{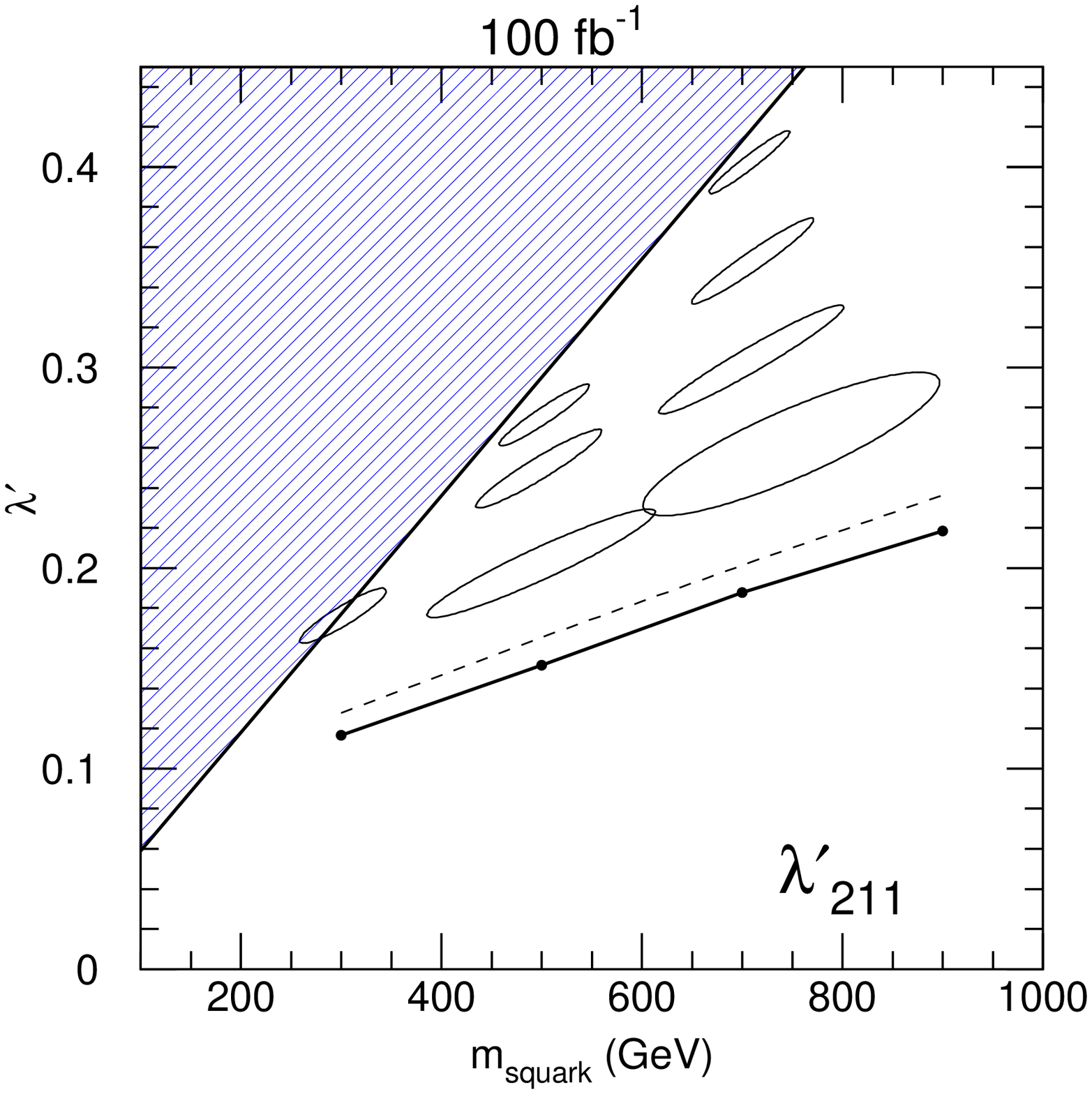}}
\vspace{-1.1cm}
\caption{\em $1 \sigma$ ellipses for the measurement of $\lambda^\prime$
and $m_{\tilde q}$ if no previous information on $m_{\tilde q}$ 
is assumed for different points in the $m_{\tilde q}$--$\lambda^\prime$
plane and an integrated luminosity of 100~pb$^{-1}$. The full and dashed lines 
correspond to the 95\% exclusion region. The analysis is performed at 
the parton level. 
The shaded region on the left is excluded
by low energy measurements.
}
\label{Fig:ell}
}

Finally, we consider the combined precision, in the measurement
of $m_{\tilde q}$ and $\lambda^\prime$, that may be achieved
 if a two-dimensional likelihood is used, and no prior
information on the squark mass is assumed. We have performed 
this study for a few sample points, and the results are 
shown, in Figure~\ref{Fig:ell},
as $1 \sigma$  ellipses in the ($m_{\tilde q}$--$\lambda^\prime$) 
plane, calculated on a sample of $\sim 500$ Monte Carlo experiments. 
The resolution in mass is of the order of few tens of GeV, and quickly 
degrades when $\lambda'$  approaches  the edge of 
the sensitivity region. Given the correlation between the $\lambda^\prime$ and
$m_{\tilde q}$ measurements, the $\lambda^\prime$ measurement is degraded 
accordingly.  This statement is better quantified in 
Table~1, 
where we give the parameters of the ellipses corresponding 
to $\log({\cal L}_{\rm max})-1/2$ for the studied points, compared with the 
resolution on $\lambda^\prime$ if the squark mass is known. In particular, 
the parameter $\rho$, which measures the correlation of the two variables, 
is $\sim$ 1, meaning that the variables are fully correlated.
Thus, a combined measurement of squark mass and $\lambda^\prime$ is 
possible with this analysis for a significant fraction of the 
parameter space. 
\begin{table}
\begin{center}
\begin{tabular}{|l|c|c|c|c|c|}
\hline
$m_{\tilde q}$ (GeV) & $\lambda^\prime$ & $\sigma(\lambda^\prime)(1)$ & $\sigma(\lambda^\prime)(2)$ & $\sigma(m_{\tilde q})$ (GeV) & $\rho$ \\
\hline
300 &  0.175 &  0.005 & 0.014 &   43 & 0.913 \\
500 &  0.200 &  0.010 & 0.027 &  114 & 0.947 \\
500 &  0.250 &  0.007 & 0.020 &   63 & 0.948 \\
500 &  0.275 &  0.006 & 0.017 &   46 & 0.951 \\
500 &  0.300 &  0.005 & 0.014 &   38 & 0.949 \\
700 &  0.250 &  0.013 & 0.036 &  148 & 0.859 \\
700 &  0.300 &  0.009 & 0.028 &   98 & 0.951 \\
700 &  0.350 &  0.007 & 0.020 &   56 & 0.954 \\
700 &  0.400 &  0.005 & 0.016 &   40 & 0.964 \\
\hline
\end{tabular}
\end{center}
\label{tab:ell}
\caption{\em Statistical errors on the $\lambda^\prime$ measurement 
for a few sample points.  The value in the third column 
[$\sigma(\lambda^\prime)(1)$] is the uncertainty 
assuming that the squark mass is known with zero error. The last three 
columns give the 
parameters of the $1 \sigma$ ellipses for the two-dimensional
likelihood on ($m_{\tilde q}$--$\lambda^\prime$), where $\rho$ is the
correlation coefficient for $m_{\tilde q}$ and $\lambda^\prime$.}
\end{table}

\section{Analysis of fully generated events}
The analysis in the previous section was performed assuming 
that the momentum of the $\mu^+\mu^-$ system has no component transverse
to the beam axis, and that the muon momenta are perfectly measured 
in the detector.

In order to perform a more realistic evaluation of the precision of 
$\lambda^\prime$ measurement, achievable in a real experiment, 
the matrix elements for the squark exchange process 
have been inserted into the PYTHIA \cite{PYTHIA}
event generator as an external process, and full events have been generated, 
including the full PYTHIA machinery for QCD showering from the initial-state 
quarks, and for the hadronization.  The events thus generated have been passed 
through the fast simulation of the ATLAS detector \cite{ATLFAST}, 
including, in particular, a very detailed parametrization of the resolution of 
the muon momentum measurement.

%
\vspace{-0.4cm}
\FIGURE{
\centerline{
\dofigs{0.42\textwidth}{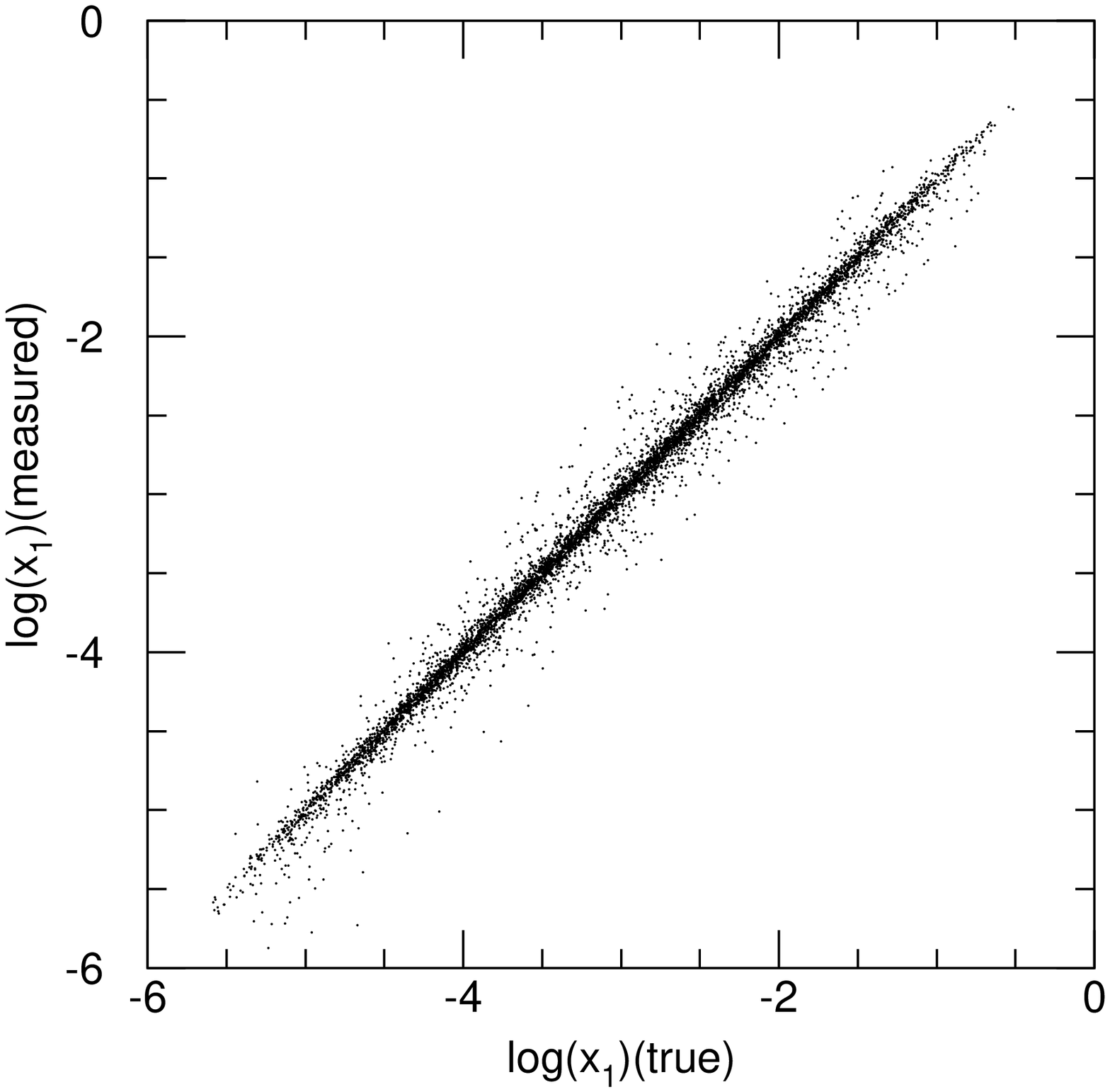}{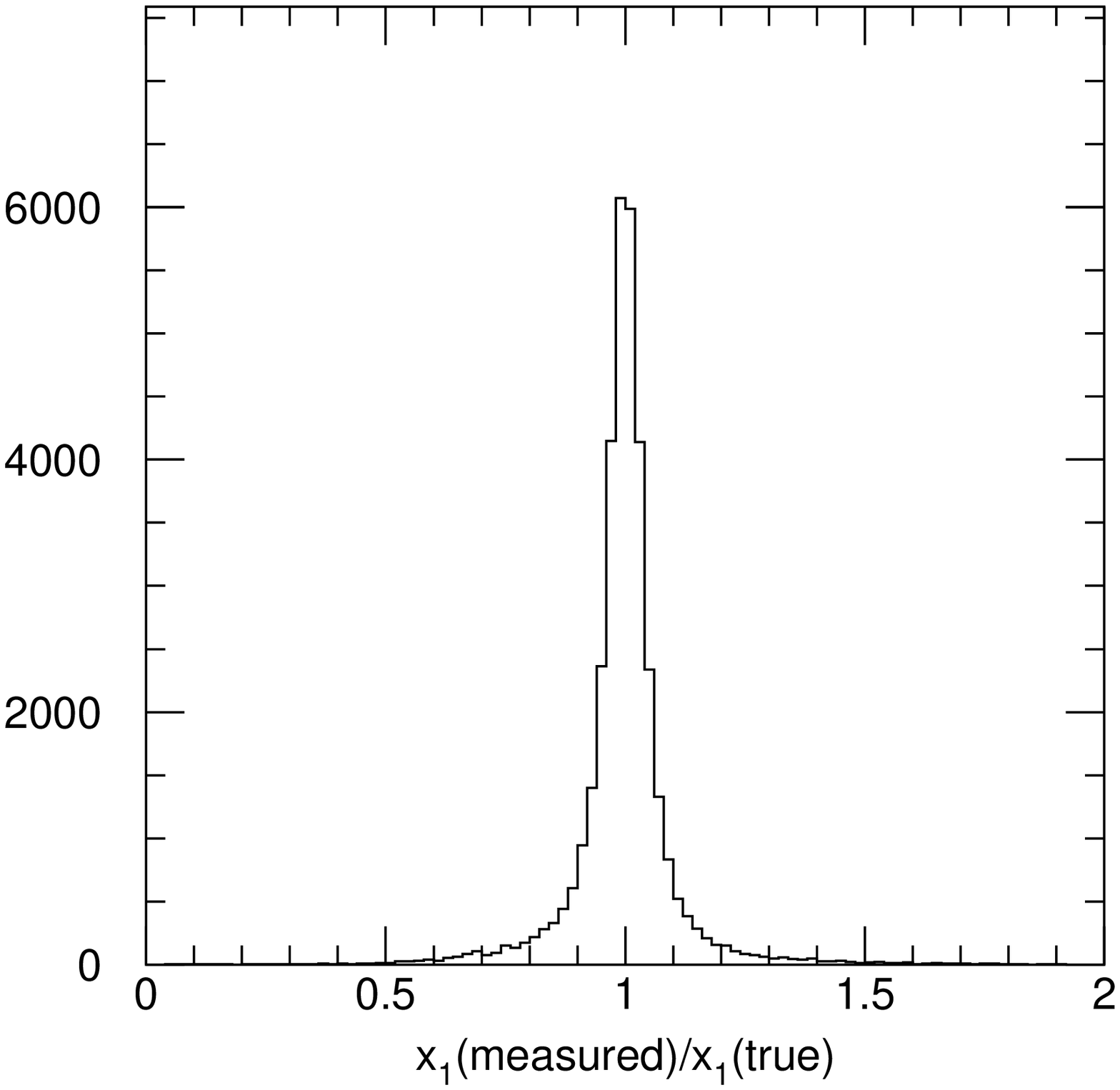}}
\vspace{-1.0cm}
\caption{\em Left plot: correlation of the reconstructed 
        value of $\log (x_1)$ 
with the generated value for a sample of events passing the analysis
requirements. Right plot: distribution of the ratio of the reconstructed 
and generated value of $x_1$. The distributions are given for a sample of
events passing the analysis cuts. }
\label{Fig:x1}
}
%
The values of $x_1$, $x_2$ have been inferred
from the measured four-momenta of the detected muons, 
according to the formulae:
$$
\frac{2\,P_L^{\mu^+\mu^-}}{\sqrt{s}}\, = \, x_1\, - \, x_2, \; \; \; 
m^2_{\mu^+\mu^-}=x_1x_2s \ .
$$
For the evaluation of $\cos\theta$ we use the Collins--Soper 
convention~\cite{CS}, which consists of an equal sharing of the $\mu^+\mu^-$ 
system transverse momentum between the two quarks.
We show, in Figure~\ref{Fig:x1}, the two-dimensional correlation 
between the generated and  the reconstructed values for the variable  $x_1$ 
as well as their ratio. Only events that pass 
the analysis criteria enter these plots. For $x_1$, the RMS deviation 
is $\sim 10\%$, and is
completely determined by the muon momentum resolution.  
A comparable resolution is obtained for $|\cos\theta|$ 
(Figure~\ref{Fig:costh}), but with significant 
tails due to the presence of a non-zero transverse momentum of the $\mu^+\mu^-$
system.
%
\vspace{-0.42cm}
\FIGURE[h]{
\centerline{
\dofigs{0.4\textwidth}{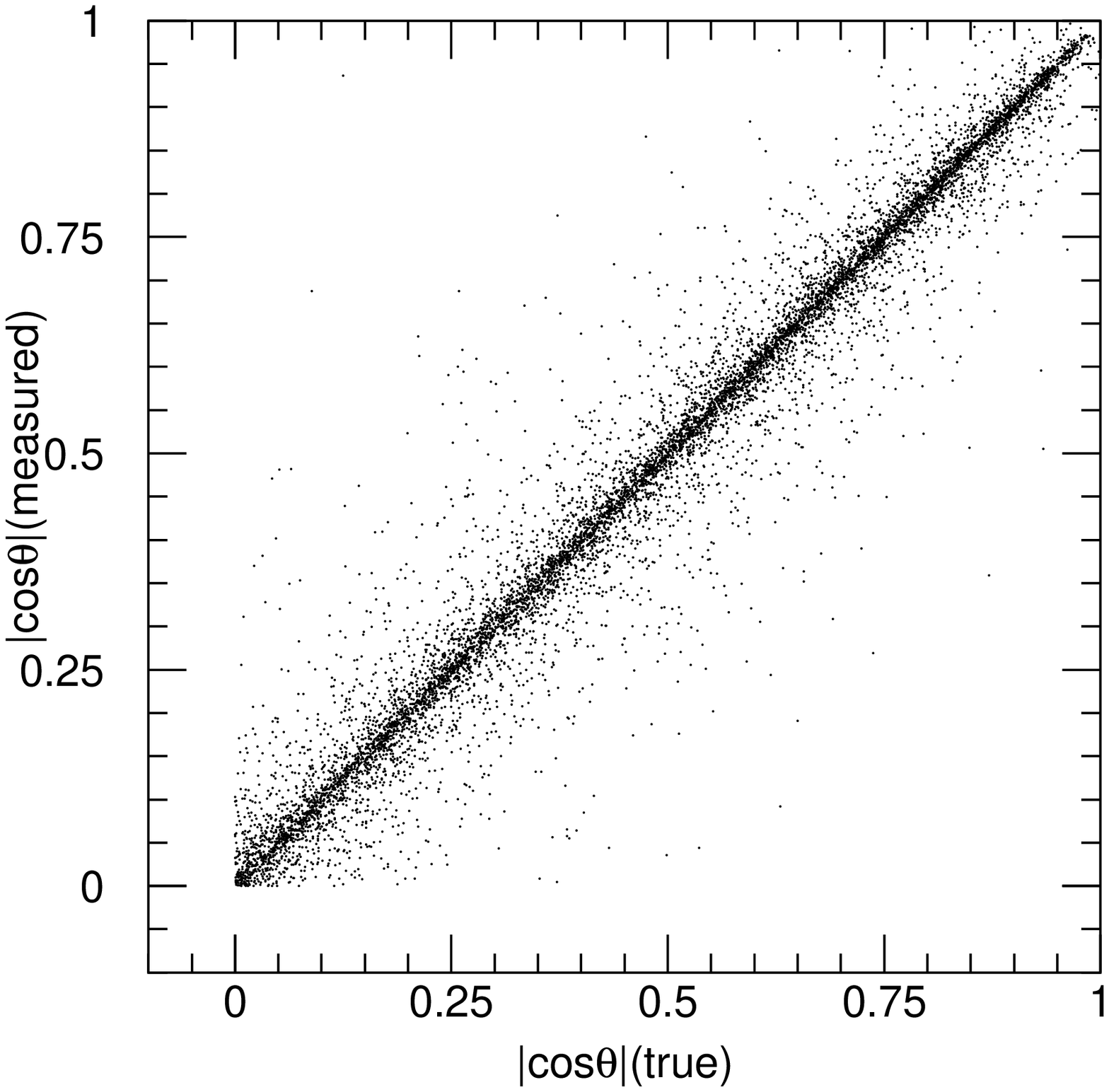}{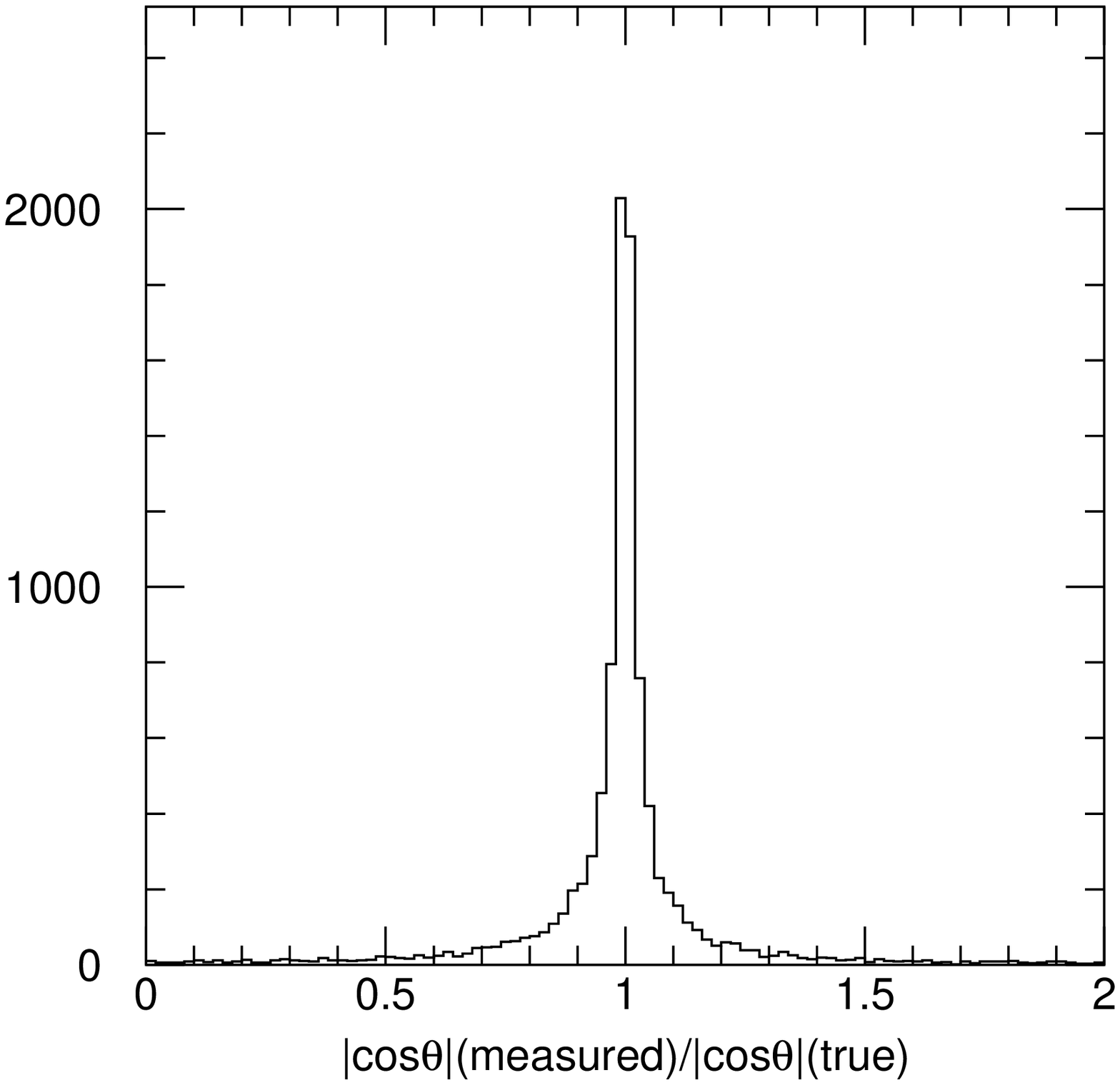}}
\vspace{-1.2cm}
\caption{\em As in Figure~\protect\ref{Fig:x1}, but for 
    $|\cos \theta|$.}
\label{Fig:costh}
}
%

Samples of events were generated for 
four squark masses, 300, 500, 700 and 900~GeV, and for three values of 
$\lambda^\prime$  for each mass, chosen in each case in an 
interval of 0.1 in $\l^{\prime 2}$, approximately covering the transition 
region where this analysis becomes sensitive to the RPV 
contribution for an integrated luminosity of 100~fb$^{-1}$, as described in 
the previous section.  We then applied to these samples the unbinned 
likelihood analysis, as for the parton-level events. The
\def\topfraction{1.}
\def\textfraction{0.}
\FIGURE[!h]{
\centerline{
\dofig{0.9\textwidth}{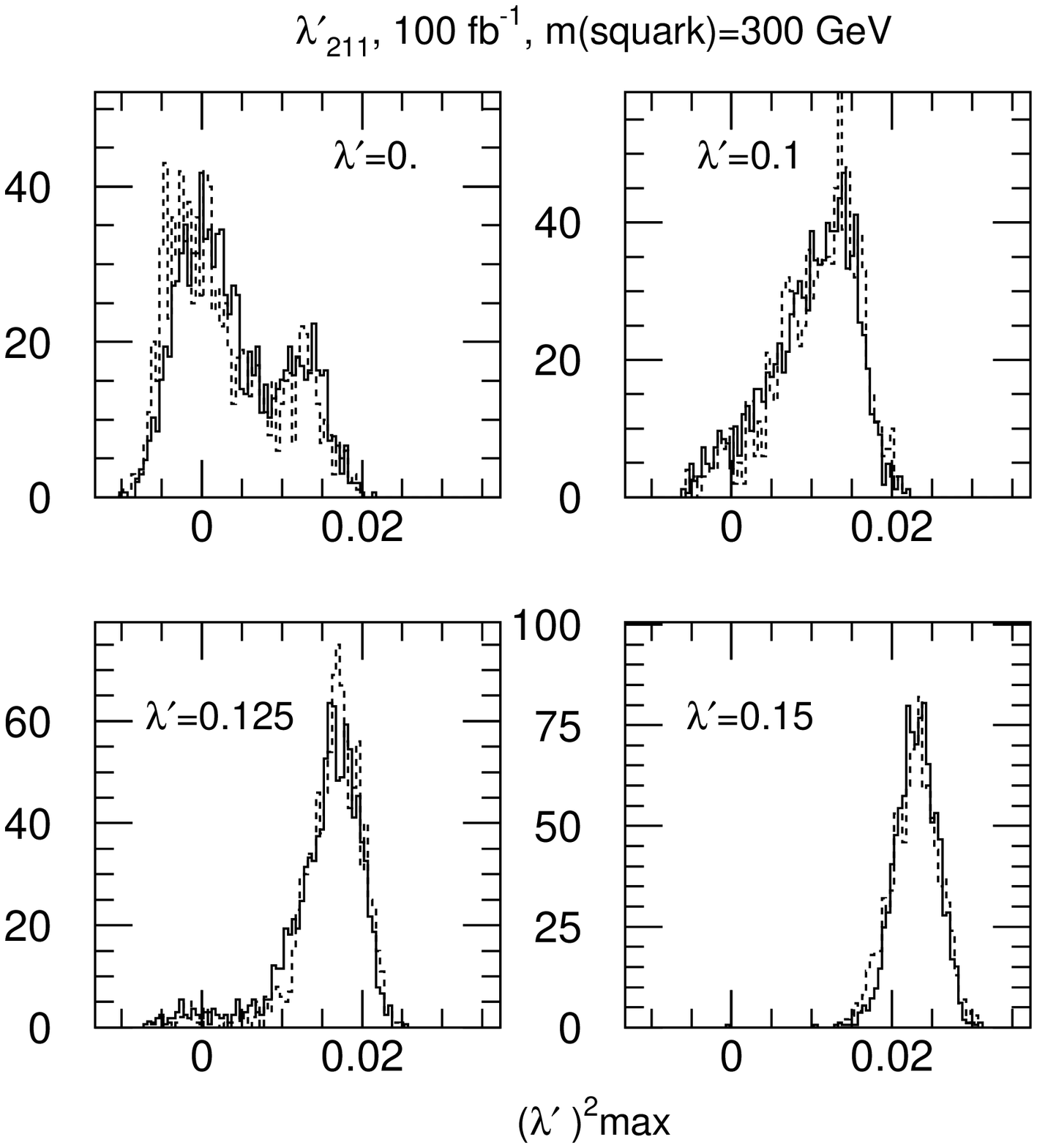}}
\vspace{-1.2cm}
\caption{\em Distributions of measured $\l^{\prime 2}$ for a set of 
Monte Carlo experiments for different values
of $\lambda^\prime$ used for the generation of events. 
The full-line histograms are the parton-level distributions
while the dashed histograms 
correspond to fully generated events. }
\label{Fig:plsme300}
}
%
evolution of the likelihood distributions with $\l'$ is 
nicely reproduced. In particular, the threshold behaviour of the 
sensitivity is still present. The distribution of the 
measured $\lambda^\prime$ is indeed dominated by the statistical 
performance of the maximum likelihood estimator, 
and not by the experimental resolution. The results of the analysis 
are shown in  Figure~\ref{Fig:plsme300}, for $m_{\tilde q} = 300$ GeV.
Therefore, the sensitivity limit 
calculated at parton level,  shown in Figure~\ref{Fig:plres}, applies
for the full event simulation as well. A difference is observed only
for $\lambda^\prime$ values for which the distribution of the measured 
$\lambda^\prime$ for a set of Monte Carlo  experiments is 
approximately gaussian.
In this case, the presence of the experimental smearing induces a moderate
increase in the $\lambda^\prime$ measurement resolution. 
Figure~\ref{Fig:diff100} shows a comparison of  
the achievable parton-level resolution with the 
corresponding experimental resolution.
The four continuous lines give the estimated 
parton-level resolution for the four considered masses. Superimposed 
as data points are the resolutions obtained for the experimental simulation.
A difference is observed mainly for the lower $\lambda^\prime$ values, where
the experimental simulation exhibits higher non-gaussian tails 
than the parton-level simulation, thus yielding a 10--20\% higher 
RMS deviation. 
For higher $\lambda^\prime$ values, the resolutions are
essentially identical.
We can thus conclude that the curves shown in Figure~\ref{Fig:plres} are 
only marginally affected by the experimental uncertainties. 
This result is obtained under quite pessimistic assumptions. 
In fact,  we use as a definition of  $F$ and its integral 
in Eq.~(\ref{eq:weight}) the pure leading order formulas in 
Eq.~(\ref{eq:ideal}), without any 
attempt at parametrizing the $\mu^+\mu^-$ transverse momentum distribution. 

%
\vspace{-0.4cm}
\FIGURE[!h]{
\centerline{
\dofig{0.9\textwidth}{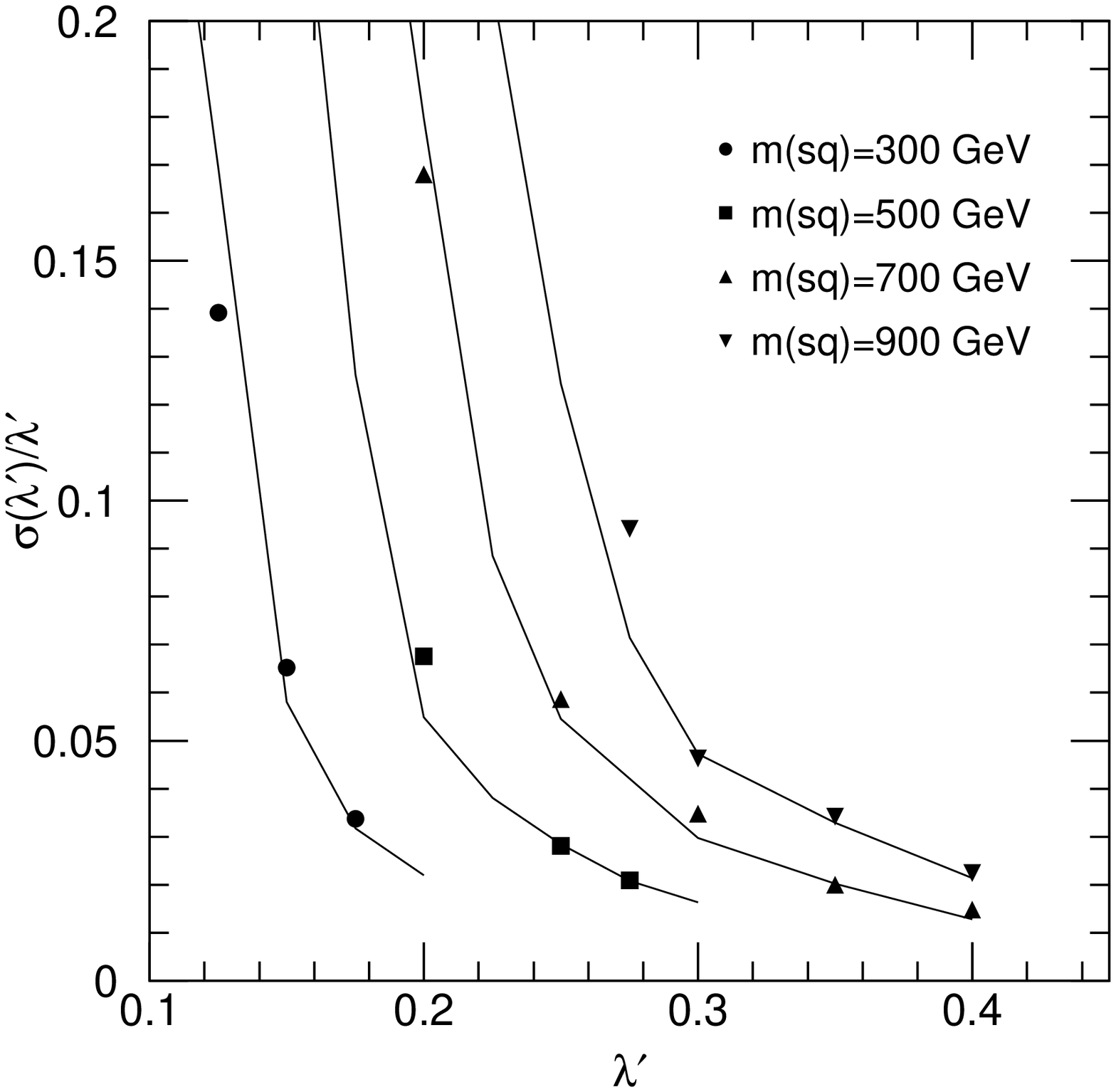}}
\vspace{-1.2cm}
\caption{\em Curves of $\sigma(\lambda')/\lambda^\prime$ as a function 
of $\lambda^\prime$ for an integrated luminosity of 100~fb$^{-1}$ and for 
four different squark masses. The full lines give the resolution 
obtained at parton level, and the points are the results of the full event
simulation.} 
\label{Fig:diff100}
}
%

In the analysis of real data, we  also need to consider the backgrounds from 
non-Drell--Yan processes. The dominant background processes,  
for the invariant mass cut applied in this analysis,
will be $\bar bb$, $\bar tt$, and $WW$  production.
For $\bar bb$ production the muons are not  isolated, and we can  apply
lepton isolation, which consists of requiring an energy deposition of 
less than 10~GeV, not
associated with the lepton in a pseudorapidity--azimuth ($\eta$--$\phi$) cone
of opening $\Delta R=0.2$ around the lepton direction.

Without any additional cuts, for an integrated luminosity of 
100~fb$^{-1}$, the backgrounds from $\bar tt$ and $WW$
amount to $\sim1400$ and $\sim500$ events respectively, for a signal of
$\sim7400$ events. The $\bar tt$ background can be strongly reduced
by vetoing  $b$-jets. We assume a tagging
efficiency of 50\% for a  rejection factor of 
100 (10) on light quark (charm) jets respectively. This is appropriate 
for the high luminosity run. We also veto jets  
tagged as $b$-jets with $P_T>30$~GeV. The $\bar tt$ background is 
thus reduced to $\sim600$ events, with negligible effect on the signal. 
Both for the $\bar tt$ and $WW$ backgrounds, the events will have real 
\Etmiss\ from escaping neutrinos, 
and a further reduction can be achieved by vetoing on high \Etmiss\ 
events. This requirement, though, has a significant effect on the 
kinematics of signal events. In fact we are considering here high 
energy  muons, 
for which the error in momentum measurement induces an instrumental imbalance
in the vector sum of the lepton momenta, which grows with increasing momentum. 
Therefore the acceptance of any kinematic cut applied must be convoluted
in the test function used for the likelihood in order to obtain the correct 
result. 
A complementary approach consists in accepting the relatively high level 
of background, and incorporating the background shape into the likelihood 
function. This should be possible with high precision, since the considered 
backgrounds yield twice as many $e\mu$ events as $\mu\mu$ events, allowing
an estimate of the background level from the data themselves. 
The two approaches have different systematic uncertainties, and can be 
used in parallel, thus providing a double check on the result.

\section{Systematic uncertainties}
In the previous section, we have studied in detail  the main
sources of experimental uncertainty, coming from the imperfect
measurement of the event variables. 
Other possible sources of experimental error are
the uncertainties in the muon energy scale, the linearity for high-energy 
muons and the acceptance evaluation. The assessment of the effect of
these uncertainties would require a detailed detector simulation,
which is outside the scope of this work. We should however 
remark that the analysis described in the previous 
section displays little sensitivity to the details of the modelling
of the experimental resolution and acceptance.  In fact, in building the 
likelihood, the crucial factor is the normalization integral in the denominator
of Eq.(\ref{eq:weight}). That integral is calculated with a sharp cut 
on the generated lepton pseudorapidity and invariant mass for
events with no $P_T^{\mu^+\mu^-}$. The data selection for the particle
level analysis is instead applied on leptons that have been smeared 
according to the detector resolution. Therefore, the acceptance of the 
cuts, especially the one on $m_{\mu^+\mu^-}$ edges, is reproduced
in a very approximate way.
Notwithstanding this fact, we observe a good agreement between 
the parton-level analysis and the particle level one. This gives us 
confidence that the additional experimental uncertainty on the
$\lambda^\prime$ measurement resolution would, at worst, only 
marginally affect the results of our analysis.
More important would be the effect of a non-linearity in the 
lepton energy measurement, which could simulate a deviation from the 
invariant-mass distribution predicted for the Standard Model.
A control on the linearity at the per cent level will be needed for
this analysis. 

We also need to consider the theoretical uncertainties in the 
likelihood calculation.
The likelihood function is built by  weighting  real events 
according to a theoretical cross section formula. Any 
discrepancy between the theoretical formula employed
and reality will induce an uncertainty on the measurement of 
$\lambda^\prime$. Two main sources of uncertainty can be identified:
1) the presence of higher-order correction to the processes and 2)
the parton distribution function (PDF) for the proton.
We consider here the most important QCD higher-order corrections 
coming from radiation from initial-state quarks, which impart 
a transverse momentum to the lepton--lepton system. From the above 
discussion of the analysis, it is clear 
that the results show very little sensitivity
to the detailed modelling of the transverse momentum distribution 
of the dilepton system.  In fact, even for the fully generated events 
the likelihood was built from the pure leading order formulas,
whereas the events are generated with the full PYTHIA machinery for 
initial-state radiation. Therefore, the experimental error
quoted in the previous section implicitly includes  very 
pessimistic assumptions on our ability to model this effect.
In a real experiment, a  more realistic theoretical 
modelling will probably be used to build the likelihood.

\FIGURE[!h]{
\centerline{
\dofig{0.80\textwidth}{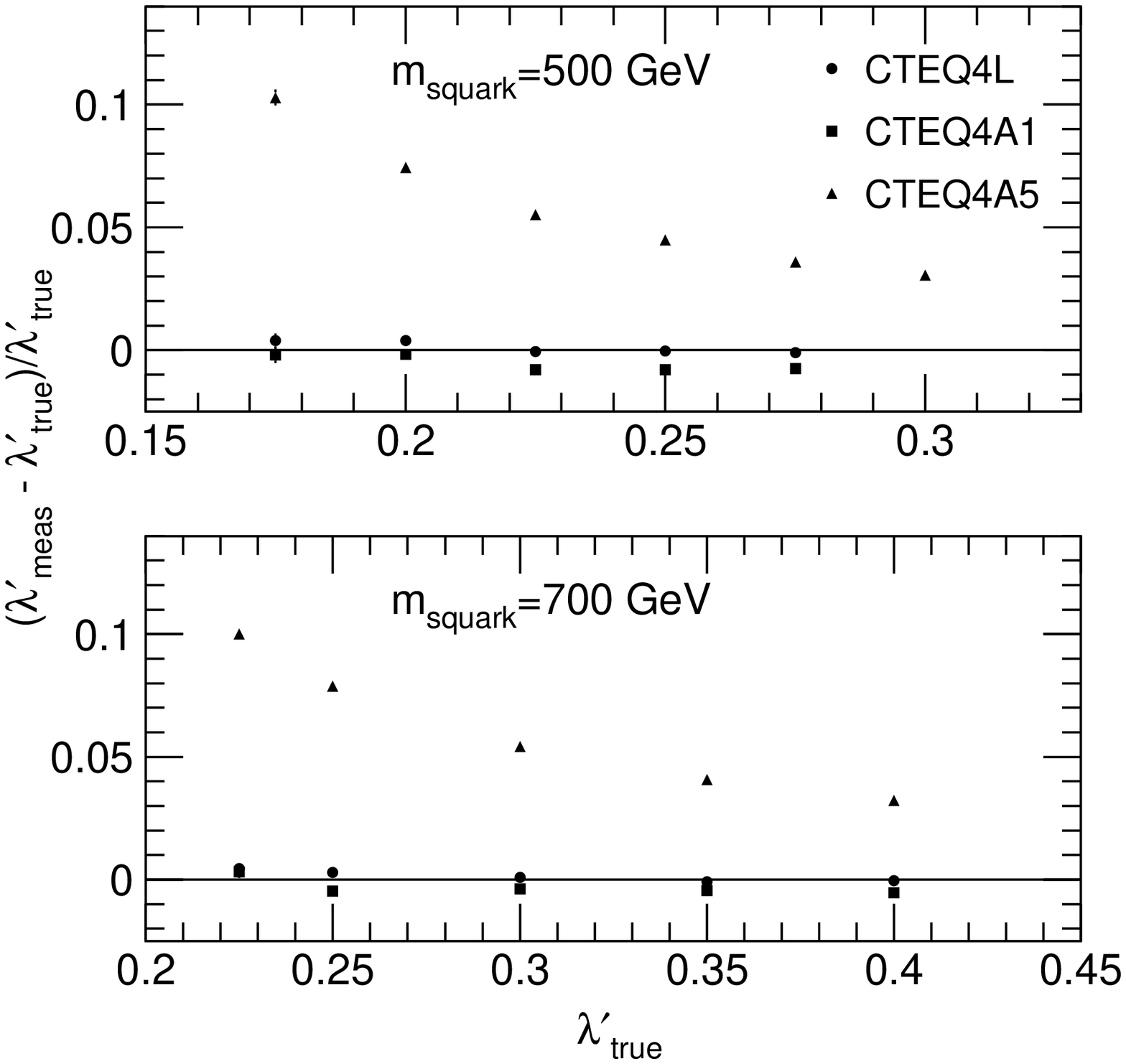}}
\vspace{-1.4cm}
\caption{\em Fractional deviation of the measured $\lambda^\prime$ 
value from the true one as a function of $\lambda^\prime$ for three different 
choices of structure functions in the calculation of the likelihood.
The events were generated with CTEQ4L. The upper plot 
is for $m_{\tilde q}=500$~GeV, the lower one for $m_{\tilde q}=700$~GeV}.
\label{Fig:plpdf}
}

The second source of uncertainty is more important, since it will affect 
the shapes of the distribution. All the events were generated with the
CTEQ4L PDF set. We have evaluated the effect of an uncertainty on 
PDF parametrization by performing the likelihood using
sets of PDFs different from the ones used for event generation. 
We chose the PDFs labelled as CTEQ4A1 and CTEQ4A5, which span an extreme
range between 213 and 399~MeV for the QCD parameter $\Lambda^{[4]}$.
The value of $\Lambda^{[4]}$ for the set CTEQ4L is 236~MeV.
We show, in Figure~\ref{Fig:plpdf}, the fractional deviation of the measured
$\lambda^\prime$ from the true one for the default structure functions 
CTEQ4L as well as for two other possible choices. 
The deviation for CTEQA1 is small, as expected from 
the small difference in $\Lambda^{[4]}$. For the set CTEQA5  the 
fractional difference
is approximately 10\% for the lowest values of $\lambda^\prime$ considered, 
and it decreases with increasing $\lambda^\prime$ to a few per cent. 
We can tentatively conclude that the uncertainty on the measurement 
coming from the choice of PDFs will be at the few per cent level, 
comparable to the statistical one.

\section{Effect of the $\rp$ couplings on the decays of the squarks}
The above discussion establishes that the dimuon production at the LHC  
has enough sensitivity to $\rp$ couplings to admit their measurement 
with a fair accuracy. Our analysis shows that a $\rp$
coupling of the order of the gauge coupling can be measured quite accurately
even for rather large squark masses, up to $700$--$900$ GeV.  Clearly,
squarks in this mass range have a substantial production cross section at
the LHC. So this naturally brings us to the question of whether one could detect
or even measure such couplings in the decays of these squarks. A discussion of
this issue is clearer if we review a few facts about the possibilities of
measuring the $\rp$ couplings in collider experiments in general. These 
are, of course, strongly dependent on the value of the $\rp$ coupling being 
probed. In the worst-case scenario, it may be so small that the only effect
it has is to cause the decay of the LSP and that too outside the detector,
so that it is  mistaken for a stable non-interacting particle. 
In this situation, we will altogether miss the phenomenon of R-parity
violation at the colliders.  For intermediate values 
of these couplings, one will get to a region where a significant fraction of
the LSPs will decay inside the inner cavity of the LHC detector and in that
case one might be able to measure the strength of these couplings by detecting
the displaced vertices inside the detector. To the best of our knowledge, a  
detailed experimental study to map the parameter space in this case is not 
currently available. Then comes the range of couplings that are large enough 
to cause a prompt decay of the LSP, giving rise to striking final states,
but are still not large enough to cause single-sparticle production or 
significantly affect decays of 
sparticles other than the LSP.  In this range of values the distinctive 
final states caused by the decay of the LSP due to the $\rp$ couplings can  
provide evidence for the existence of $\rp$, but cannot give any information 
whatsoever on their size.  And finally comes the region of the even 
larger $\rp$ couplings that we have considered. Here,
 one is sensitive to the effects
of these couplings via virtual exchanges of sparticles on scattering processes 
as well as decays of sparticles, other than the LSP, caused by them.
We have demonstrated  in the above work the feasibility of  `measuring' 
such a $\rp$ coupling through the contribution of virtual squark exchanges to 
the dimuon production. This offers perhaps the cleanest way of measuring the 
$\rp$ couplings with an adequate control of the experimental and theoretical 
systematic uncertainties, as the only input required for this is the squark 
mass.  As already mentioned, for values of the squark masses under 
consideration, their production cross sections at the LHC are substantial. 
Hence, the effect of the $\rp$ 
couplings on the decays of the squark produced via the strong interactions 
needs to be studied. A particularly attractive case will be when the $\rp$
decay of the squark competes with an $R_p$-conserving decay with an 
identifiable signature. A comparison of the two can then give a direct 
measurement of the $\rp$ coupling. A promising example for the couplings
we have considered is the $\rp$ decay $\tilde t \rightarrow \mu + $ jet. 
To study this issue, we will need to consider both the total production cross 
sections and the branching fractions into the various channels. 

\FIGURE[!h]{
\centerline{
\epsfxsize=7.5cm\epsfysize=10.0cm
                     \epsfbox{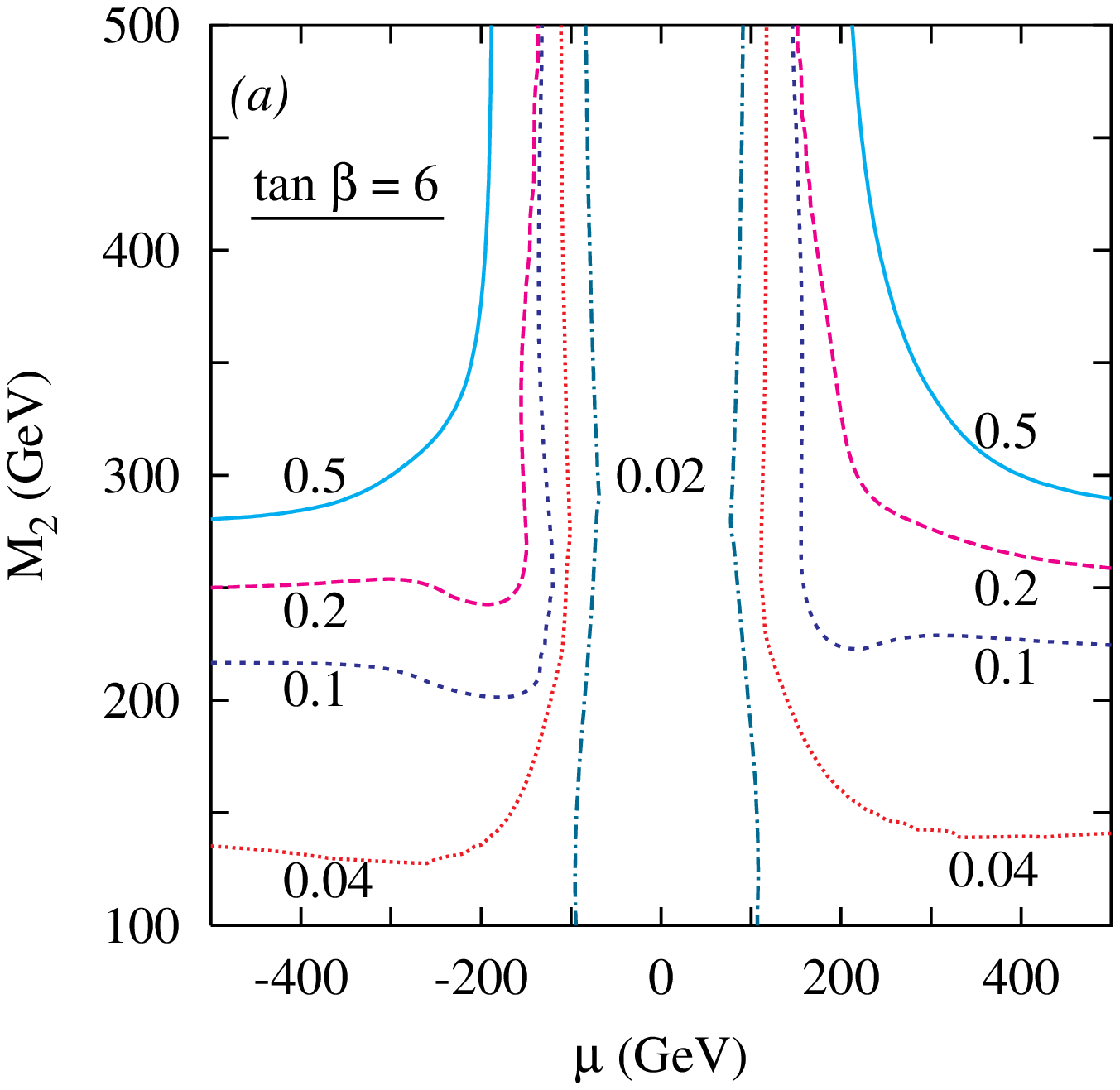}
        \hspace*{1ex}
\epsfxsize=7.5cm\epsfysize=10.0cm
                     \epsfbox{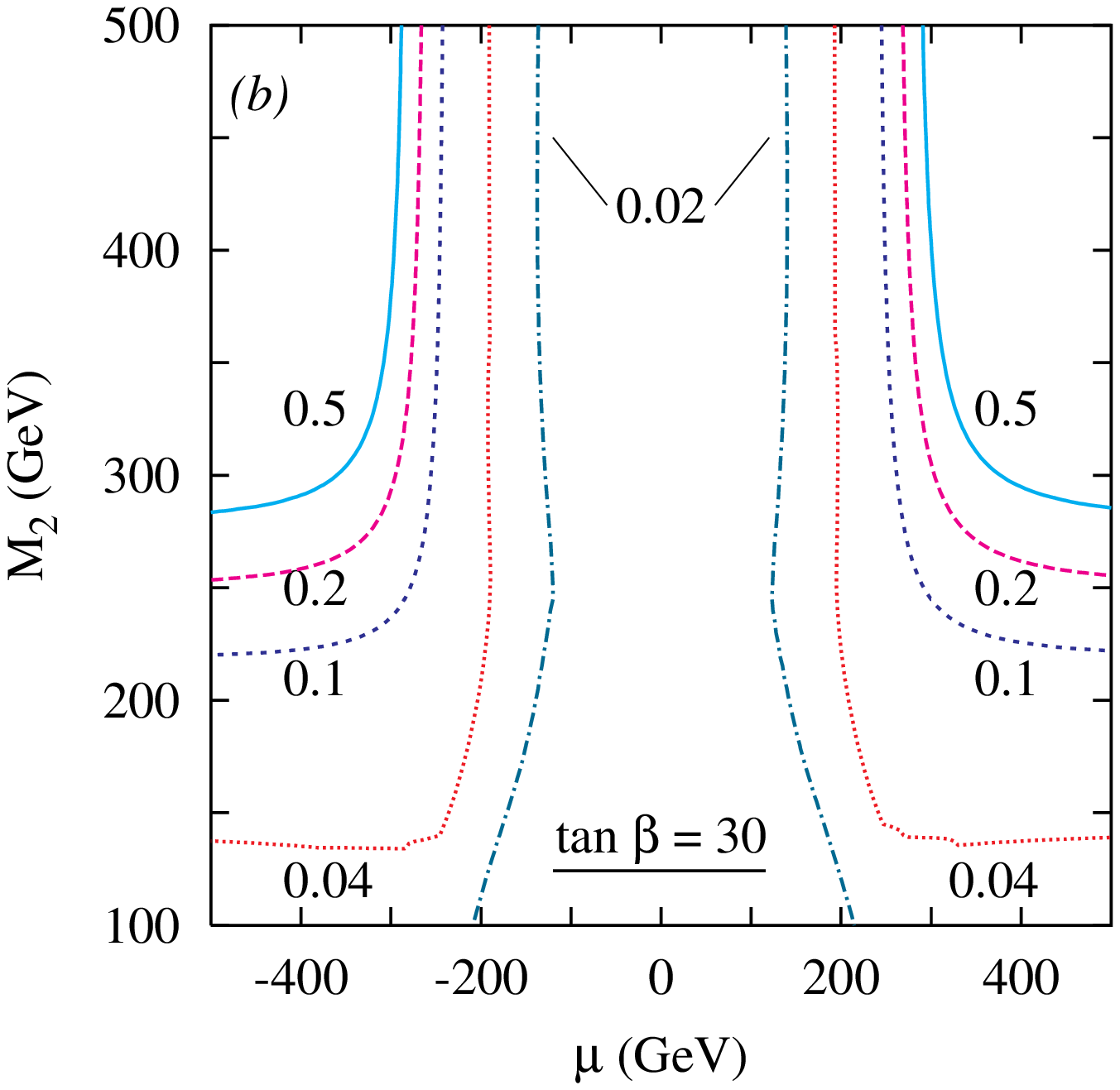}
}
\vspace*{-15ex}
\caption{\em Contours for the branching fraction of a $300 \gev$ stop 
        ($\tilde t_L$) into the $R$-parity-violating channel. 
        The unification relation between the gaugino masses 
        has been assumed.  The value of the $R$-violating coupling 
        ($\lambda'_{231} = 0.16$) is the minimum necessary  for a 
        determination  $\lambda'$ to an accuracy of $5\% $
        (see Figure~\protect\ref{Fig:plres}). The left and right panels
        correspond to two different values of $\tan \beta$.
        }
\label{Fig:br_300}
}

Squark pair-production has been analysed at length in various 
papers and depends crucially on the gluino mass.  A crude summary of
these results relevant to our discussion is that
``for any squark lighter than approximately 800 GeV, 
the pair-production cross section at the LHC is 
large enough to lead to discovery''.  
For the most part, the reach in mass and possibility of
discovery will be independent of whether $R$-parity is violated or not. 
The measurement of the relevant coupling, or even the 
establishment of its identity, would, however, require that 
at least two channels be measurable with a relatively high degree 
of accuracy. One of these should be  $R$-parity-violating,  
while the other should be  $R$-conserving.
For top-squarks, the presence of leptons and $b$-jets in the final 
state provides a handle for the identification of the exclusive
decays of interest. 
The relevant $R$-conserving decay channels are 
$\tilde q \rightarrow q \tilde g, q \tilde \chi^0_j, 
                      q' \tilde \chi^\pm_j, \tilde q' W$. 
Each of these channels could lead to a cascade process 
culminating in a $\rp$ decay. We are, however, concerned 
with the $\rp$ decay of the original squark.
In Figure~\ref{Fig:br_300}, we display $\rp$-branching 
fraction contours (into $\mu^+ + d$) 
for the decay of a 300 GeV stop. The assumed 
value of $\l'_{231}$ is the least that would permit a 
detection with a statistical accuracy of 5\%, if any dilution in the 
efficiency of the analysis from surviving backgrounds is disregarded. 
We see that barring very light charginos, the $\rp$ mode tends to dominate. 
An overwhelmingly large  $\rp$ branching fraction, while leading to 
spectacular signals, is hardly amenable to a precision measurement of the 
coupling. Similar arguments hold for a very small $\rp$ branching fraction. 
The best hope is therefore for moderate values of 
$M_2$ and $\mu$. It is thus clear that, at the level 
of branching ratios, stop decays would provide 
a sufficient number of events for the measurement
of the $\rp$ coupling over a significant parameter space. 
A detailed experimental analysis, outside the scope
of this work,  is however needed 
to ascertain whether it will be possible to isolate the
final states produced by the two relevant processes, 
with adequate efficiency, and with a level of purity that
would permit a reasonably accurate determination of the relevant branching 
fractions.

\section{Conclusions}
To summarize, we demonstrate that an analysis of the Drell--Yan process
at the LHC detectors would be a significant tool in the task of probing
the parameter space in an $R_p$-violating supersymmetric model. While the
deviation in the total cross section is at best a few per cent, even for
large values of the relevant $R_p$-violating coupling constant, the
differential distributions are much more discriminating, the distribution
in the dilepton invariant mass proving particularly useful.  By adopting
the maximum likelihood  method we could maximize the sensitivity 
of the measurement, avoiding at the same time the normalization 
uncertainties due to structure functions as well as higher-order corrections, 
etc.

Working, for definiteness, with dimuon production in the ATLAS detector,
we showed that, for a wide range of the parameter space, $R_p$-violating
supersymmetry would be amenable to discovery through this process. Even
more importantly, a measurement of the squark mass as well as the
$\rp$ coupling would be possible with relatively low errors: 2--3\% for the
coupling and somewhat larger for the mass. These errors could be reduced even 
further if additional information about the squark mass were to become
available from other measurements. We found the systematic errors 
to be small. The analysis proved to be  quite robust and the results
gleaned from fully generated events were very similar to those 
obtained from a parton-level study.  Since for the range of squark masses 
to  which  
our investigation is sensitive,  they can be pair-produced copiously at the 
LHC, we also further  looked into the possibility of determining {\it the
same} $\rp$ coupling, for which we analysed the sensitvity using the  
Drell--Yan process,  through the decays of the squarks.

Finally, let us add that even though  we have confined ourselves to dimuon 
production in a particular theory (namely $R_p$-violating supersymmetry), 
a similar analysis could be carried out for dielectron production 
equally well.  
In general such an analysis can be used to study effects of any alternate
theories of physics, beyond the SM, which affect the Drell--Yan process.
Our analysis demonstrates that such studies would complement very well the 
more direct methods for new particle search.

\bigskip 

\acknowledgments

This work was initiated during a workshop held in Les Houches. We warmly thank
Patrick Aurenche and all of the organizing team for the stimulating
program, the excellent atmosphere and the outstanding computing
facilities.  We thank the ATLAS collaboration members for useful discussions.
We have made use of the physics analysis framework and
tools, which are the result of collaboration-wide efforts.
DC thanks the Theory Divsion, CERN 
for hospitality while part of the project was being carried out
and the Dept. of Science and Technology, India, for financial assistance 
under the Swarnajayanti Fellowship grant.

\end{document}